\documentclass[twocolumn,showpacs,aps,floatfix,prd,nofootinbib]{revtex4-1}

\usepackage[utf8]{inputenc}
\usepackage{graphicx}
\usepackage{dcolumn}
\usepackage{array}
\usepackage{longtable}
\usepackage{amsmath}
\usepackage{mathtools}
\usepackage[per-mode=symbol,separate-uncertainty=true]{siunitx}
\usepackage{nicefrac}
\usepackage{afterpage}

\newcommand{\BaBarNumber}     {PUB-17/002}
\newcommand{\SLACPubNumber}   {SLAC-PUB-17147}

\long\def\inst#1{\par\nobreak\kern 4pt\nobreak
    {\it #1}\par\vskip 10pt plus 3pt minus 3pt}

\DeclareSIUnit\clight{\text{\ensuremath{c}}}
\DeclareSIUnit\MeVperc{\MeV\per\clight}
\DeclareSIUnit\MeVpercsq{\MeV\per\clight\squared}
\DeclareSIUnit\GeVpercsq{\GeV\per\clight\squared}

\RequirePackage{xspace}

\usepackage{relsize}
\def\babar{\mbox{\slshape B\kern-0.1em{\smaller A}\kern-0.1em
    B\kern-0.1em{\smaller A\kern-0.2em R}}\xspace}

\def\babartit{\mbox{B\kern-0.1em{\smaller A}\kern-0.1em
    B\kern-0.1em{\smaller A\kern-0.2em R}}\xspace}

\def\Kbar  {\kern 0.2em\overline{\kern -0.2em K}{}\xspace}

\def\Kz    {\ensuremath{K^0}\xspace}
\def\Kzb   {\ensuremath{\Kbar^0}\xspace}
\def\KzKzb {\ensuremath{\Kz \kern -0.16em \Kzb}\xspace}
\def\Kp    {\ensuremath{K^+}\xspace}
\def\Km    {\ensuremath{K^-}\xspace}

\def\KpKm  {\ensuremath{\Kp \kern -0.16em \Km}\xspace}

\def\Dbar    {\kern 0.2em\overline{\kern -0.2em D}{}\xspace}

\def\Dz      {\ensuremath{D^0}\xspace}
\def\Dzb     {\ensuremath{\Dbar^0}\xspace}
\def\DzDzb   {\ensuremath{\Dz {\kern -0.16em \Dzb}}\xspace}
\def\Dp      {\ensuremath{D^+}\xspace}
\def\Dm      {\ensuremath{D^-}\xspace}

\def\DpDm    {\ensuremath{\Dp {\kern -0.16em \Dm}}\xspace}

\def\Bbar    {\kern 0.18em\overline{\kern -0.18em B}{}\xspace}

\def\Bz      {\ensuremath{B^0}\xspace}
\def\Bzb     {\ensuremath{\Bbar^0}\xspace}
\def\BzBzb   {\ensuremath{\Bz {\kern -0.16em \Bzb}}\xspace}
\def\Bu      {\ensuremath{B^+}\xspace}
\def\Bub     {\ensuremath{B^-}\xspace}

\def\BpBm    {\ensuremath{\Bu {\kern -0.16em \Bub}}\xspace}

\def\BorBbar    {\kern 0.18em\optbar{\kern -0.18em B}{}\xspace}
\def\DorDbar    {\kern 0.18em\optbar{\kern -0.18em D}{}\xspace}
\def\KorKbar    {\kern 0.18em\optbar{\kern -0.18em K}{}\xspace}

\def\jpsi     {\ensuremath{{\uppercase{J}\mskip -3mu/\mskip -2mu\psi\mskip 2mu}}\xspace}
\def\psitwos  {\ensuremath{\psi{(2S)}}\xspace}

\mathchardef\Upsilon="7107
\def\Y#1S{\ensuremath{\Upsilon{(#1S)}}\xspace}

\def\FourS {\Y4S}

\mathchardef\Deltares="7101
\mathchardef\Xi="7104
\mathchardef\Lambda="7103
\mathchardef\Sigma="7106
\mathchardef\Omega="710A

\def\Deltabar{\kern 0.25em\overline{\kern -0.25em \Deltares}{}\xspace}
\def\Lbar{\kern 0.2em\overline{\kern -0.2em\Lambda\kern 0.05em}\kern-0.05em{}\xspace}
\def\Sigbar{\kern 0.2em\overline{\kern -0.2em \Sigma}{}\xspace}
\def\Xibar{\kern 0.2em\overline{\kern -0.2em \Xi}{}\xspace}
\def\Obar{\kern 0.2em\overline{\kern -0.2em \Omega}{}\xspace}
\def\Nbar{\kern 0.2em\overline{\kern -0.2em N}{}\xspace}
\def\Xb{\kern 0.2em\overline{\kern -0.2em X}{}\xspace}

\def\cm   {\ensuremath{{\rm \,cm}}\xspace}

\def\barn{\ensuremath{{\rm \,b}}\xspace}

\def\mus  {\ensuremath{\rm \,\mus}\xspace}

\def\mus        {\ensuremath{\,\mu{\rm s}}\xspace}    

\def\rad{\ensuremath{\rm \,rad}\xspace}

\def\to                 {\ensuremath{\rightarrow}\xspace}

\def\pep2{PEP-II\xspace}

\def\gsim{{~\raise.15em\hbox{$>$}\kern-.85em
          \lower.35em\hbox{$\sim$}~}\xspace}
\def\lsim{{~\raise.15em\hbox{$<$}\kern-.85em
          \lower.35em\hbox{$\sim$}~}\xspace}

\xspace

\def\jetset74   {\mbox{\tt Jetset \hspace{-0.5em}7.\hspace{-0.2em}4}\xspace}

\newcommand{\defeq}{\mathrel{\mathop:}=}

\makeatletter
\newcommand{\colorcaption}[2][]{%
  \renewcommand{\@caption@fignum@sep}{ (color online). }%
  \caption[#1]{#2}%
}
\makeatother

\newcommand\amunew{\num{17.9}}
\newcommand\amunewstaterr{\num{0.1}}
\newcommand\amunewsysterr{\num{0.6}}
\newcommand\amunewtotrelerr{\SI{3.3}{\percent}}

\newcommand\amuhinew{\num{17.4}}
\newcommand\amuhinewstaterr{\num{0.1}}
\newcommand\amuhinewsysterr{\num{0.6}}
\newcommand\amuhinewtotrelerr{\SI{3.2}{\percent}}

\newcommand\amuwidenew{\num{21.8}}
\newcommand\amuwidenewstaterr{\num{0.1}}
\newcommand\amuwidenewsysterr{\num{0.7}}

\newcommand\alphanew{\num{4.44}}
\newcommand\alphanewstaterr{\num{0.02}}
\newcommand\alphanewsysterr{\num{0.14}}

\newcommand\alphawidenew{\num{6.58}}
\newcommand\alphawidenewstaterr{\num{0.02}}
\newcommand\alphawidenewsysterr{\num{0.22}}

\newcommand\systpeak{\SI{3.1}{\percent}}
\newcommand\systjpsi{\SI{6.7}{\percent}}
\newcommand\systhigh{\SI{7.2}{\percent}}

\newcommand\errpeak{\SI{3.6}{\percent}}

\newcommand\threshul{\SI{1.2}{\GeVpercsq}}

\begin{document}

\begin{flushleft}
  {\babar}-\BaBarNumber \\
  \SLACPubNumber \\
  \vspace{1.4mm}
\end{flushleft}

\title{
\large \bf \boldmath
Measurement of the $e^+e^-\to\pi^+\pi^-\pi^0\pi^0$ cross section using initial-state radiation at \babar
}

\author{J.~P.~Lees}
\author{V.~Poireau}
\author{V.~Tisserand}
\affiliation{Laboratoire d'Annecy-le-Vieux de Physique des Particules (LAPP), Universit\'e de Savoie, CNRS/IN2P3,  F-74941 Annecy-Le-Vieux, France}
\author{E.~Grauges}
\affiliation{Universitat de Barcelona, Facultat de Fisica, Departament ECM, E-08028 Barcelona, Spain }
\author{A.~Palano}
\affiliation{INFN Sezione di Bari and Dipartimento di Fisica, Universit\`a di Bari, I-70126 Bari, Italy }
\author{G.~Eigen}
\affiliation{University of Bergen, Institute of Physics, N-5007 Bergen, Norway }
\author{D.~N.~Brown}
\author{Yu.~G.~Kolomensky}
\affiliation{Lawrence Berkeley National Laboratory and University of California, Berkeley, California 94720, USA }
\author{M.~Fritsch}
\author{H.~Koch}
\author{T.~Schroeder}
\affiliation{Ruhr Universit\"at Bochum, Institut f\"ur Experimentalphysik 1, D-44780 Bochum, Germany }
% ...
\collaboration{The \babar\ Collaboration}
\noaffiliation


\begin{abstract}
The process $e^+e^- \to \pi^+\pi^-2\pi^0\gamma$ is investigated by means of the initial-state radiation technique, where a photon is emitted from the incoming electron or positron. Using \SI{454.3}{\femto\barn^{-1}} of data collected around a center-of-mass energy of $\sqrt{s} = \SI{10.58}{\GeV}$ by the \babar experiment at SLAC, approximately \num{150000} signal events are obtained. The corresponding non-radiative cross section is measured with a relative uncertainty of \errpeak\ in the energy region around \SI{1.5}{\GeV}, surpassing all existing measurements in precision. Using this new result, the channel's contribution to the leading order hadronic vacuum polarization contribution to the anomalous magnetic moment of the muon is calculated as
$(g_\mu^{\pi^+\pi^-2\pi^0}-2)/2 = (\amunew \pm \amunewstaterr_\mathrm{stat} \pm \amunewsysterr_\mathrm{syst}) \times 10^{-10}$ in the energy range $\SI{0.85}{\GeV} < E_\mathrm{CM} < \SI{1.8}{\GeV}$.
In the same energy range, the impact on the running of the fine structure constant at the $Z^0$-pole is determined as
$\Delta\alpha^{\pi^+\pi^-2\pi^0}(M^2_\mathrm{Z}) = (\alphanew \pm \alphanewstaterr_\mathrm{stat} \pm \alphanewsysterr_\mathrm{syst}) \times 10^{-4}$.
Furthermore, intermediate resonances are studied and especially the cross section of the process $e^+e^- \to \omega\pi^0 \to \pi^+\pi^-2\pi^0$ is measured.
\end{abstract}


\maketitle

\setcounter{footnote}{0}


\section{Introduction}
\label{sec:Introduction}

The anomalous magnetic moment of the muon, $g_\mu-2$, exhibits a discrepancy of more than three standard deviations~\cite{Davier:2010nc} between experiment and theory, making it one of the most interesting puzzles in contemporary particle physics.
New experiments to improve the measurement of $g_\mu-2$ are starting operation at Fermilab~\cite{Chapelain:2017syu} and J-PARC~\cite{Kitamura:2017xyx}.
On the theoretical side~\cite{Passera:2004bj}, the QED and weak contributions account for the largest contribution to $g_\mu-2$ and have been calculated with precision significantly exceeding the experiment. The theoretical prediction is limited by the hadronic contributions, which cannot be calculated perturbatively at low energies. Therefore, measured cross sections are used in combination with the optical theorem to compute the hadronic part of $g_\mu-2$. This leads to the dominant uncertainty in the standard model prediction of $g_\mu-2$, which is comparable to the experimental precision. Hence, in order to improve the theoretical prediction, accurate measurements of all hadronic final states are needed.
In this paper, we present a new measurement of one of the least known cross sections, $e^+e^- \to \pi^+\pi^-2\pi^0$.
This measurement supersedes a preliminary analysis~\cite{Druzhinin:2007cs} from \babar on the same final state. The earlier measurement was performed on approximately half of the \babar data set. Additionally, the new analysis improves the systematic uncertainties of the detection efficiency and of the background subtraction.

The limited precision of this cross section also limits the precision of the running of the fine structure constant $\Delta\alpha$. 

The \babar experiment is operated at fixed center-of-mass (CM) energies in the vicinity of \SI{10.58}{\GeV}. Therefore, the method of initial-state radiation (ISR) is used to determine a cross section over a wide energy range. This method uses events where one of the initial particles radiates a photon, thus lowering the effective CM-energy available for hadron production in the electron-positron annihilation process.
Events where the photon is emitted as final-state radiation (FSR) can be neglected since their produced number is extremely low and the FSR photon rarely is sufficiently energetic. Hence, the resulting \emph{radiative} cross section is then converted back into the \emph{non-radiative} cross section using the relation~\cite{Bonneau:1971mk}
\begin{equation}
\frac{\mathrm{d}\sigma_{\pi^+\pi^-2\pi^0\gamma}(M)}{\mathrm{d}M} = \frac{2M}{s} \cdot W(s,x,C) \cdot \sigma_{\pi^+\pi^-2\pi^0}(M) \text{.}
\end{equation}
The radiative cross section of the final state $\pi^+\pi^-2\pi^0$ is denoted by $\sigma_{\pi^+\pi^-2\pi^0\gamma}$, while $\sigma_{\pi^+\pi^-2\pi^0}$ is the non-radiative equivalent.
The variable $s$ is the square of the CM energy of the experiment, $x = \frac{2 E^\ast_\gamma}{\sqrt{s}}$, $E^\ast_\gamma$ is the CM energy of the ISR photon, and $M = \sqrt{(1-x)s}$ the invariant mass of the hadronic final state, equivalent to the effective CM energy $E_\mathrm{CM}$ of the hadronic system.
The radiator function $W(s,x,C)$ describes the probability at the squared CM energy $s$ for an ISR photon of energy $E^\ast_\gamma$ to be emitted in the polar angle range $| \cos\theta^\ast_\gamma | < C$.
It is calculated to leading order in a closed form expression~\cite{Czyz:2000wh}, while next-to-leading order effects are accounted for by simulation using PHOKHARA~\cite{Czyz:2008kw,Czyz:2010hj}.

This paper is structured as follows: in Sec.~\ref{sec:babar}, the \babar detector and the analyzed data set are described. Section~\ref{sec:sel} outlines the basic event selection and the kinematic fit, while Sec.~\ref{sec:bkg} illustrates the background removal procedure. Acceptance and efficiency determination are explained in Sec.~\ref{sec:eff}. The main results -- cross section and contributions to $a_\mu \defeq (g_\mu - 2)/2$ as well as $\Delta\alpha$ -- are presented in Sec.~\ref{sec:cs}, followed by the investigation of intermediate resonances in Sec.~\ref{sec:intstr}.


\section{The \babar detector and data set}
\label{sec:babar}

The \babar experiment was operated at the \pep2 storage ring at the SLAC National Accelerator Laboratory. Its CM energy was mainly set to the \FourS resonance at \SI{10.58}{\GeV}, while smaller samples were taken at other energies. In this analysis, the full data set around the \FourS is used, amounting to an integrated luminosity of \SI{454.3}{\femto\barn^{-1}}~\cite{Lees:2013rw}. The \babar detector is described in detail elsewhere~\cite{Aubert:2001tu,TheBABAR:2013jta}. 
The innermost part of the detector is a silicon vertex tracker (SVT), surrounded by the Drift Chamber (DCH), both operating in a \SI{1.5}{\tesla} magnetic field. Together, the SVT and DCH provide tracking information for charged particles. Neutral particles and electrons are detected in the electromagnetic calorimeter (EMC), which also measures their energy. Particle identification (PID) is provided by the information from the EMC, SVT, and DCH combined with measurements from the internally reflecting ring-imaging Cherenkov detector (DIRC). Muons are identified using information from the instrumented flux return (IFR) of the solenoid magnet, consisting of iron plates interleaved with resistive plate chambers and, in the later runs, limited streamer tubes.

The detector response to a given final state is determined by a detector simulation based on GEANT4~\cite{Agostinelli:2002hh}, which accounts for changes in the experimental setup over time.

Using the AfkQed~\cite{Bevan:2014iga} event generator, based on EVA~\cite{Binner:1999bt,Czyz:2000wh}, simulation samples of ISR channels are produced. These include the signal process (for efficiency calculation) as well as the background channels $\pi^+\pi^-\pi^0\gamma$, $2(\pi^+\pi^-\pi^0)\gamma$, $K^+K^-2\pi^0\gamma$, and $K_\mathrm{s}K^\pm\pi^\mp\gamma$.
For the reaction $e^+e^- \to \pi^+\pi^-3\pi^0\gamma$ two simulations exist within AfkQed, which differ by the presence of the intermediate resonances. The simulated processes are $e^+e^- \to \omega2\pi^0\gamma$ (with $\omega \to \pi^+\pi^-\pi^0$) and $e^+e^- \to \eta\pi^+\pi^-\gamma$ (with $\eta \to 3\pi^0$).
An $e^+e^- \to \tau^+\tau^-$ sample was generated with KK2f~\cite{Jadach:1999vf}.
In addition, the JETSET~\cite{Sjostrand:1993yb} generator is used to obtain a sample of continuum $e^+e^- \to q\bar{q}$ events (\textit{uds}-sample) to investigate non-ISR-background contributions in data.

PHOKHARA~\cite{Czyz:2008kw,Czyz:2010hj}, an event generator for ISR processes that includes the full NLO matrix elements, is used to cross-check the signal simulation and account for next-to-leading order ISR. 
Final-state radiation is simulated using PHOTOS~\cite{Barberio:1993qi}.

\section{Event Selection and Kinematic Fit}\label{sec:sel}

For the final state $\pi^+\pi^-2\pi^0\gamma$ two charged tracks and at least five photons must be detected, since only the decay $\pi^0 \to 2\gamma$ is  considered. The photon of highest CM energy is chosen as the ISR photon and is required to have an energy of at least \SI{3}{\GeV}. Furthermore, it must lie in the laboratory frame polar angle range $\SI{0.35}{\rad} < \theta_\mathrm{ISR} < \SI{2.4}{\rad}$, in which detection efficiencies have been extensively studied~\cite{Bevan:2014iga}.
The distance of closest approach of a charged track to the beam axis in the transverse plane is required to be less than \SI{1.5}{\cm}. The distance of the point closest to the beam axis is required to be less than \SI{2.5}{\cm} along the beam-axis from the event vertex.
Additionally, the tracks are restricted to the polar angle range $\SI{0.4}{\rad} < \theta_\mathrm{tr} < \SI{2.45}{\rad}$ in the laboratory frame and must have a transverse momentum of at least \SI{100}{\MeVperc}. In order to select the back-to-back topology typical for ISR events with a hard photon, the minimum laboratory frame angle between the ISR photon and a charged track has to exceed \SI{1.2}{\rad}.

Photons with an energy in the laboratory frame $E_{\gamma\mathrm{lab}} > \SI{50}{\MeV}$ and with a polar angle within the same range as the ISR photon are considered to build the $\pi^0$ candidates (the charged track vertex is assumed as their point of origin). The invariant mass of each two-photon combination is required to be within \SI{30}{\MeVpercsq} of the nominal $\pi^0$ mass~\cite{PDG}, while the resolution is about \SI{7}{\MeVpercsq}. An event candidate is then built with the two selected tracks, the ISR photon, and any pair of $\pi^0$ candidates with no photons in common, with the further requirement that at least one of the four photons has to have an energy $E_{\gamma\mathrm{lab}} > \SI{100}{\MeV}$.

\begin{figure}
  \includegraphics[width=\linewidth]{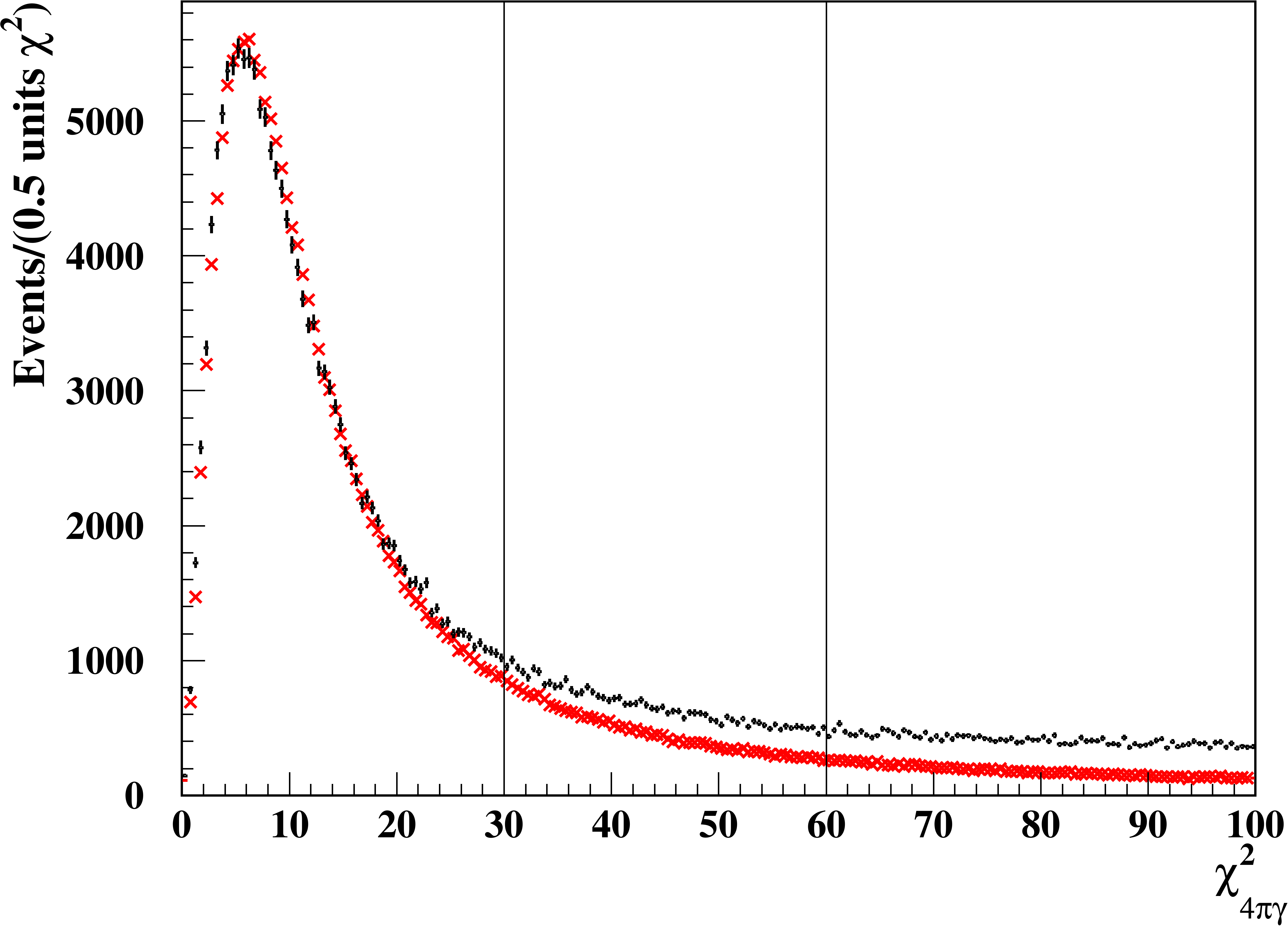}
  \colorcaption{The $\chi^2_{\pi^+\pi^-2\pi^0\gamma}$ distributions after full selection for data (black points) and the AfkQed generator (red crosses, normalized to the same area as data in the range $\chi^2_{\pi^+\pi^-2\pi^0\gamma} < 10$). The vertical lines indicate the signal ($\chi^2_{\pi^+\pi^-2\pi^0\gamma} < 30$) and sideband ($30 < \chi^2_{\pi^+\pi^-2\pi^0\gamma} < 60$) regions used for background subtraction.}
  \label{fig:chi2add}
\end{figure}

Candidate events are subjected to a kinematic fit in the hypothesis $e^+e^- \to \pi^+\pi^-2\pi^0\gamma$ with six constraints (four from energy-momentum conservation and two from the $\pi^0$ mass). 
The photon combination achieving the best fit result is subsequently used in the reconstructed event.
The distribution of $\chi^2_{\pi^+\pi^-2\pi^0\gamma}$, the $\chi^2$ of the kinematic fit, is shown in Fig.~\ref{fig:chi2add} for data and simulation after full selection (also including the selection criteria described in the following paragraphs).
The latter distribution is normalized to data in the region $\chi^2_{\pi^+\pi^-2\pi^0\gamma} < 10$, where a lower background level is expected. The $\chi^2_{\pi^+\pi^-2\pi^0\gamma}$ distributions in data and in the AfkQed simulation sample are similar in shape, but the tail of the data distribution shows the presence of background processes, which are discussed in Sec.~\ref{sec:bkg}.
Only events with $\chi^2_{\pi^+\pi^-2\pi^0\gamma} < 30$ are selected. 
    
Besides the kinematic fit to the signal hypothesis, the events are subjected to kinematic fits of the background hypotheses $e^+e^- \to \pi^+\pi^-3\pi^0\gamma$, $e^+e^- \to \pi^+\pi^-\pi^0\gamma$, $e^+e^- \to \pi^+\pi^-\pi^0\eta\gamma$, and $e^+e^- \to \pi^+\pi^-2\eta\gamma$ if the detected number of photons is sufficient for the respective hypothesis. As in the signal hypothesis, the photon pairs are constrained to the mass of the $\pi^0$ or $\eta$ meson in the kinematic fit. The same criteria are applied to the photons as well as to the mass of each two-photon combination as in the kinematic fit to the signal hypothesis (replacing the nominal $\pi^0$ mass by the $\eta$ mass where applicable), and the best combination is selected. In the latter three hypotheses above, the resulting $\chi^2$ values are used to reject the corresponding background channels. The contribution from $\pi^+\pi^-\pi^0\gamma$ is suppressed by imposing the requirement $\chi^2_{\pi^+\pi^-\pi^0\gamma} \geq 25$. The possible background channels $\pi^+\pi^-\pi^0\eta\gamma$ and $\pi^+\pi^-2\eta\gamma$ (with $\eta \to 2\gamma$ in both cases) are rejected through the requirements $\chi^2_{\pi^+\pi^-\pi^0\eta\gamma} > \chi^2_{\pi^+\pi^-2\pi^0\gamma}$ and $\chi^2_{\pi^+\pi^-2\eta\gamma} > \chi^2_{\pi^+\pi^-2\pi^0\gamma}$. The background from $e^+e^- \to \pi^+\pi^-3\pi^0\gamma$ is removed as outlined in Sec.~\ref{ssec:bkg5pi}.

Events containing kaons or muons are suppressed by using the \babar PID-algorithms as outlined in Sec.~\ref{ssec:bkgkaon} and in Sec.~\ref{ssec:bkgmuon}, respectively.

\section{Background}\label{sec:bkg}
\begin{figure}
  \centering
  \includegraphics[type=pdf,ext=.pdf,read=.pdf, width=\linewidth]{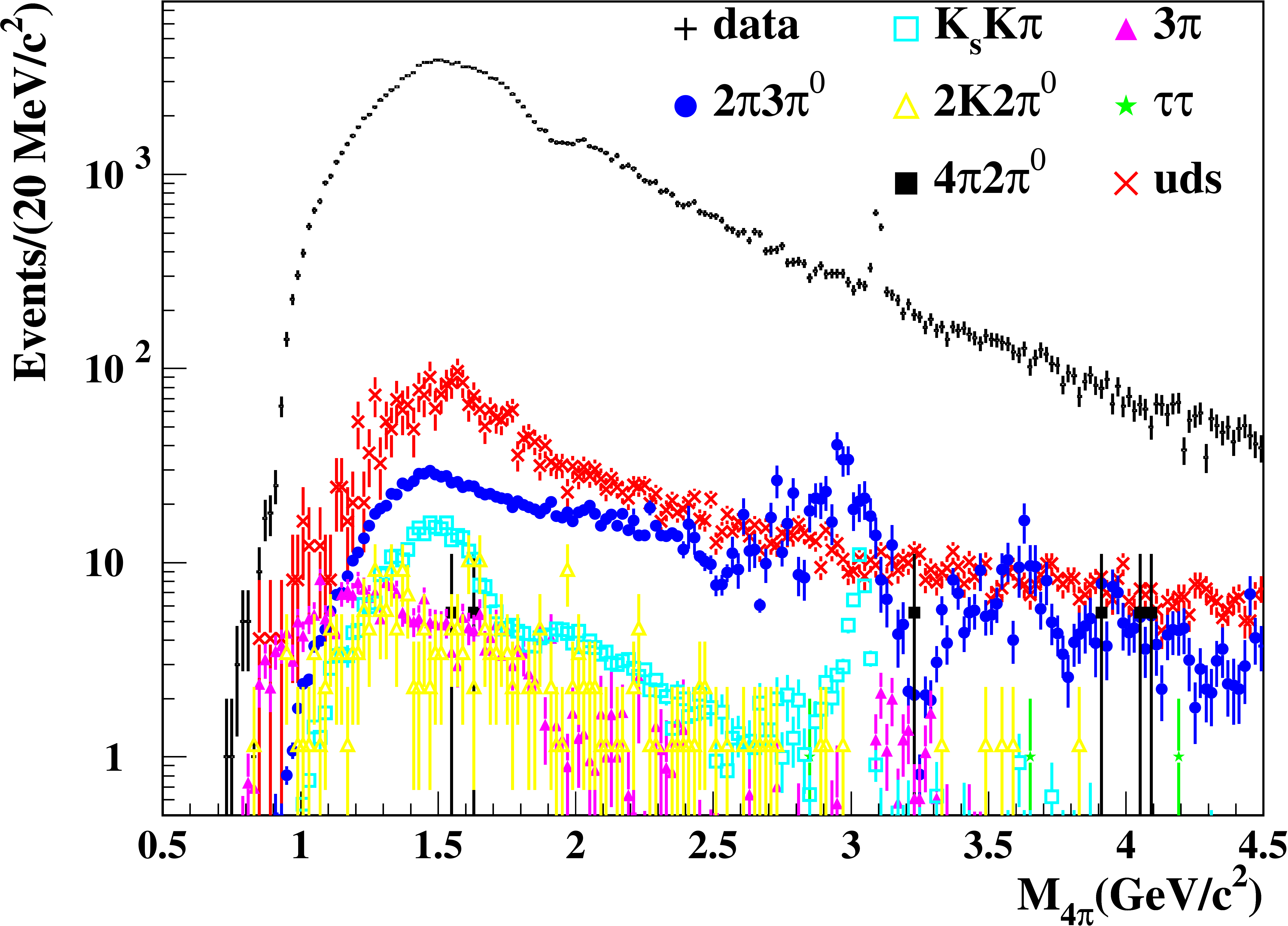}
  \colorcaption{The $\pi^+\pi^-2\pi^0\gamma$ data (black points) compared to simulated backgrounds after selection: \textit{uds}-continuum (red crosses), $\pi^+\pi^-3\pi^0\gamma$ (blue solid circles), $K_\mathrm{s}K^\pm\pi^\mp\gamma$ (turquoise open squares), $K^+K^-2\pi^0\gamma$ (yellow open triangles), $\pi^+\pi^-\pi^0\gamma$ (pink solid triangles), $2(\pi^+\pi^-\pi^0)\gamma$ (black solid squares), and $\tau^+\tau^-$ (green solid stars) as a function of $M(\pi^+\pi^-2\pi^0)$.}
  \label{fig:allbkg}
\end{figure}

Background events originate from continuum hadron production, hadron production via ISR, and the leptonic channel $e^+e^- \to \tau^+\tau^-$, all shown in Fig.~\ref{fig:allbkg}.
Most events from such processes are removed by the selection outlined above, but specific vetoes are needed for particular channels containing kaons or muons, the latter predominantly produced in the decay $e^+e^- \to \jpsi2\pi^0\gamma \to \mu^+\mu^-2\pi^0\gamma$. Furthermore, remaining background events are subtracted using simulation and sideband subtraction.
The channel $e^+e^- \to \pi^+\pi^-3\pi^0\gamma$ was determined to be the largest ISR background contribution. Since this process has not been measured with sufficient precision before, it is treated separately in a dedicated measurement reported below.

\subsection{Continuum Processes}
The largest background contribution originates from continuum hadron production. 
 In order to subtract this contribution, a simulation based on the JETSET generator~\cite{Sjostrand:1993yb} is used after modifications discussed below to make it more precise.
The \textit{uds}-MC events including a true photon (e.g., ISR or FSR photon, but not a photon from, e.g., a $\pi^0$-decay) with $E_\gamma > \SI{3}{\GeV}$ at generator level are discarded.
As the remaining continuum MC-sample does not contain ISR events, a photon from a $\pi^0$ decay must be misidentified as an ISR photon for the event to pass the selection criteria.
Since the relative fraction of low-multiplicity events in the continuum simulation is rather unreliable, the continuum sample is normalized by comparing the $\pi^0$ peak in the invariant $\gamma_\mathrm{ISR}\gamma$ mass to data (considering all $\gamma_\mathrm{ISR}\gamma$ combinations, where $\gamma_\mathrm{ISR}$ is the selected ISR photon and $\gamma$ corresponds to any photon not already assigned to a $\pi^0$). The normalization scales the number of continuum events down by approximately a factor of three compared to the prediction by the generator (with a relative uncertainty of the normalization of roughly \SI{20}{\percent}) and is applied as a function of the invariant mass $M(\pi^+\pi^-2\pi^0)$ to give a precise result over the full energy range.
 As is visible in Fig.~\ref{fig:allbkg}, continuum processes, which are subtracted using simulation, amount to approximately \SI{3}{\percent} of data in the peak region.

\subsection{\boldmath$e^+e^- \to \pi^+\pi^-3\pi^0\gamma$}\label{ssec:bkg5pi}

\begin{figure}
  \centering
  \includegraphics[width=\linewidth]{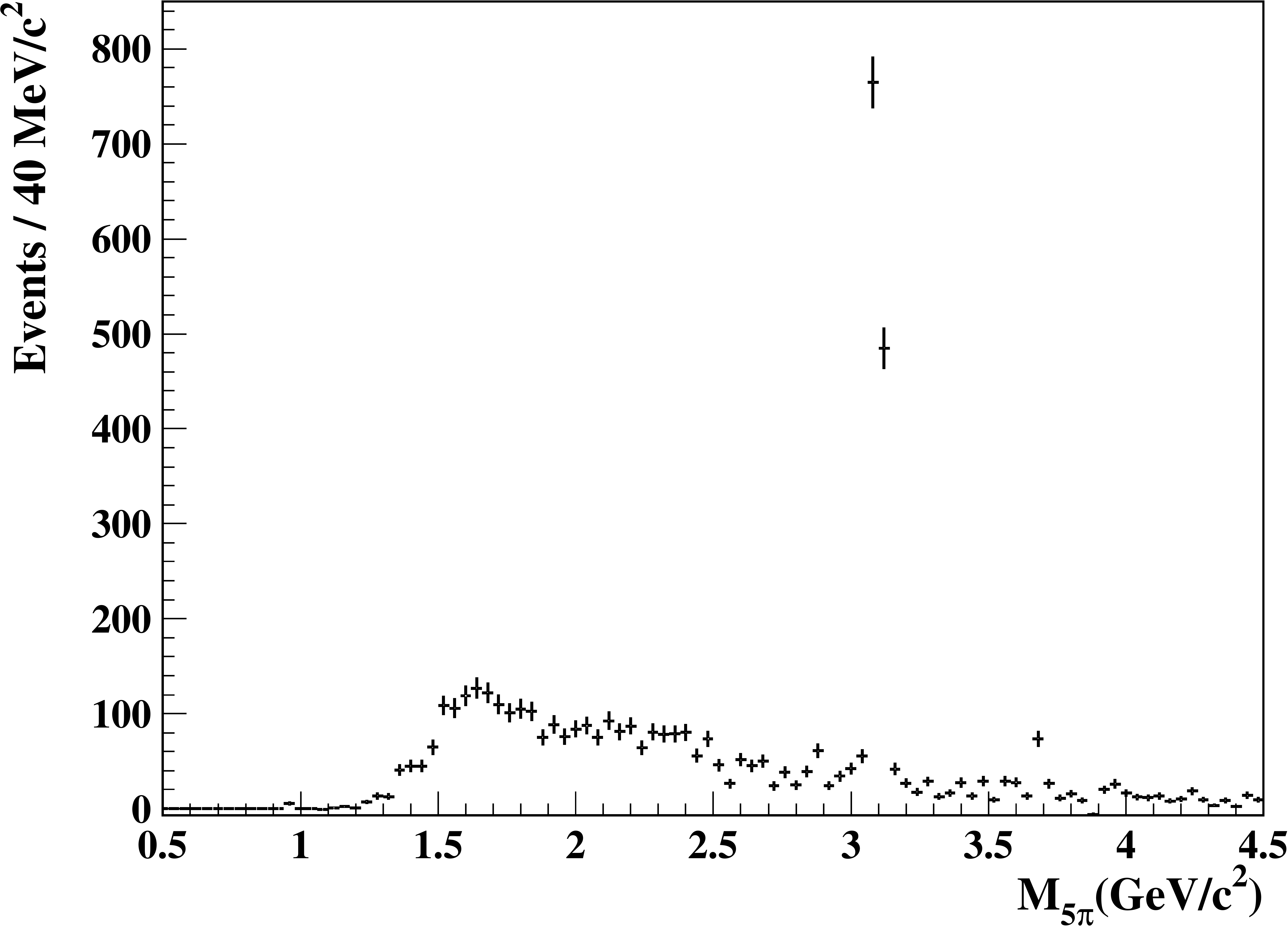}
  \caption{Measured $M(\pi^+\pi^-3\pi^0)$ distribution of the $e^+e^- \to \pi^+\pi^-3\pi^0\gamma$ background channel in data.}
  \label{fig:5pimass-raw}
\end{figure}

Since this channel has so far only been measured with large uncertainties~\cite{Cosme:1978qe}, a dedicated study was performed. For this purpose, candidate events are subjected to the kinematic fit under the hypothesis $e^+e^- \to \pi^+\pi^-3\pi^0\gamma$.
In this study continuum background is subtracted using the sample generated by JETSET, while ISR background is subtracted employing the method outlined in Sec.~\ref{ssec:novobkg}.
The resulting measured event spectrum is shown in Fig.~\ref{fig:5pimass-raw}.
The detection efficiency of $\pi^+\pi^-3\pi^0\gamma$ events is calculated using simulated samples of the intermediate states $\omega2\pi^0\gamma$ and $\eta\pi^+\pi^-\gamma$. Due to their distinct kinematics, the $\chi^2_{\pi^+\pi^-3\pi^0\gamma}$ distributions differ and hence the detection efficiencies determined from either $\omega2\pi^0\gamma$ or $\eta\pi^+\pi^-\gamma$ differ by up to \SI{67}{\percent} from each other, depending on the invariant mass $M_{\pi^+\pi^-3\pi^0}$.
 
Studying the $3\pi^0$ and $\pi^+\pi^-\pi^0$ invariant mass distributions in data, it was found that -- neglecting interference -- about \SI{38}{\percent} of the $\pi^+\pi^-3\pi^0\gamma$ events are produced via $\omega2\pi^0\gamma$ and about \SI{26}{\percent} via $\eta\pi^+\pi^-\gamma$, both for $M_{\pi^+\pi^-3\pi^0} < \SI{2.9}{\GeVpercsq}$. 
Hence, less than \SI{40}{\percent} of the $\pi^+\pi^-3\pi^0\gamma$ events are produced through other channels or phase space.
Since there is no simulation of this fraction of events, a mixture according to the measured production fractions of $\omega2\pi^0\gamma$ and of $\eta\pi^+\pi^-\gamma$ is used to estimate the detection efficiency.
It has been checked in the almost background-free data sample around the \jpsi resonance that the efficiency of the $\chi^2_{\pi^+\pi^-3\pi^0}$ requirement is in excellent agreement between data and the simulation mixture, showing relative differences of less than \SI{2}{\percent}.
The difference between the $\omega2\pi^0\gamma$ and $\eta\pi^+\pi^-\gamma$ efficiencies is taken as the uncertainty for the event fraction not simulated by the $\omega2\pi^0\gamma$ or $\eta\pi^+\pi^-\gamma$ samples.
This results in a total relative uncertainty of \SI{27}{\percent} for the $e^+e^- \to \pi^+\pi^-3\pi^0\gamma$ production rate. Other uncertainties are found to be smaller.

The $M_{\pi^+\pi^-3\pi^0}$ invariant mass distributions in the $\omega2\pi^0\gamma$ and $\eta\pi^+\pi^-\gamma$ simulations differ significantly from the measured $\pi^+\pi^-3\pi^0\gamma$ mass distribution. In order to make the simulation samples as realistic as possible and to use them to estimate the background due to $\pi^+\pi^-3\pi^0\gamma$ events in the $\pi^+\pi^-2\pi^0\gamma$ event sample, their $M_{\pi^+\pi^-3\pi^0}$ distributions are adjusted to reproduce the measured event distribution. For this purpose each MC event is weighted with the factor $N_\mathrm{measured}/N_\text{MC true}$ depending on the event mass $M_{\pi^+\pi^-3\pi^0}$, where $N_\mathrm{measured}$ is the number of events measured in data after efficiency correction and $N_\mathrm{MC true}$ is the number of events produced in simulation.
The $\pi^+\pi^-2\pi^0\gamma$ selection has different rejection rates for each simulation sample, since the $\pi^+\pi^-2\pi^0\gamma$ selection is sensitive to the kinematics of the production process. 
Therefore, the efficiencies of the $\eta\pi^+\pi^-\gamma$ and  $\omega2\pi^0\gamma$ simulation samples differ by up to \SI{50}{\percent} from the mixture of both samples. This number is taken as the uncertainty of the events not produced via $\eta\pi^+\pi^-\gamma$ or $\omega2\pi^0\gamma$, where the efficiency of the mixture is assumed.

This study shows that the $e^+e^- \to \pi^+\pi^-3\pi^0\gamma$ background channel is responsible for less than \SI{1}{\percent} of the events in the peak region $\SI{1}{\GeVpercsq} \leq M(\pi^+\pi^-2\pi^0) < \SI{1.8}{\GeVpercsq}$, less than \SI{3}{\percent} for $\SI{1.8}{\GeVpercsq} \leq M(\pi^+\pi^-2\pi^0) < \SI{2.7}{\GeVpercsq}$, and less than \SI{10}{\percent} of the events for higher masses.
It is the dominant ISR background contribution, as seen from the result in Fig.~\ref{fig:allbkg}. 

Both uncertainties outlined above need to be considered, namely the uncertainty of the $\pi^+\pi^-3\pi^0\gamma$ yield (\SI{27}{\percent}) and the uncertainty of the rejection rate of $\pi^+\pi^-3\pi^0\gamma$ events in the $\pi^+\pi^-2\pi^0\gamma$ selection (\SI{20}{\percent}). Although both uncertainties have a common source they are conservatively assumed to be independent and added in quadrature.
This results in a total relative uncertainty of \SI{33}{\percent} of the $\pi^+\pi^-3\pi^0\gamma$ background level.

Hence for $\SI{1}{\GeVpercsq} < M(\pi^+\pi^-2\pi^0) < \SI{1.8}{\GeVpercsq}$ the $e^+e^- \to \pi^+\pi^-3\pi^0\gamma$ background yields an uncertainty of less than \SI{0.33}{\percent}, \SI{1.0}{\percent} for $M(\pi^+\pi^-2\pi^0) < \SI{2.7}{\GeVpercsq}$, and \SI{3.3}{\percent} for higher masses, relative to the measured number of $\pi^+\pi^-2\pi^0$ events. As will be shown in Sec.~\ref{ssec:bkgerr}, this is consistent with the independent final estimate for the background systematics.

\subsection{Kaonic Final States}\label{ssec:bkgkaon}

Two sizable background channels including kaons exist: $e^+e^- \to K^+K^-2\pi^0\gamma$ and $e^+e^- \to K_\mathrm{s}K^\pm\pi^\mp\gamma$. These final states are suppressed by requiring none of the charged tracks to be selected as a kaon by the particle identification algorithm. This algorithm uses a likelihood-based method outlined in Ref.~\cite{Bevan:2014iga} and introduces a systematic uncertainty of $\SI{0.5}{\percent}$. As shown in Fig.~\ref{fig:allbkg}, the remaining background contributions amount to \SI{0.5}{\percent} and \SI{0.25}{\percent} for $K_\mathrm{s}K^\pm\pi^\mp\gamma$ and $K^+K^-2\pi^0\gamma$, respectively, and are subtracted via simulation.

\subsection{Muonic Final States}\label{ssec:bkgmuon}

The only sizable muon contribution is produced by the channel $e^+e^- \to \jpsi2\pi^0\gamma \to \mu^+\mu^-2\pi^0\gamma$. Therefore a combined veto is applied. If the invariant mass of the two charged tracks is compatible with the \jpsi mass and at least one of the charged tracks is identified as a muon, the event is rejected. Tracks are identified as muons using a cut-based approach combining information from the electromagnetic calorimeter and the instrumented flux return~\cite{Aubert:2001tu,TheBABAR:2013jta}. 
It is observed that this combined veto rejects up to \SI{70}{\percent} of the data sample around the $\psitwos$ mass, while its effect is negligible in the remaining mass range. Due to the uncertainty of the selector, a systematic uncertainty of \SI{2}{\percent} is introduced in the $\psitwos$ region. 

Despite the dedicated veto, a number of muon events still survives the selection due to inefficiency and misidentification of the PID algorithm. 
Since the muon identification efficiency and $\pi^\pm \to \mu^\pm$ misidentification probability are well known for the \babar PID procedures, the remaining muon contribution is calculated from the data and subsequently removed. This yields a remaining muon background at the \psitwos peak of approximately $\SI{4}{\percent}$ of the data, while the rest of the mass spectrum is negligibly affected.

After removing the muonic backgrounds, no $\psitwos$ peak is observed in data.

\subsection{Additional Background Contributions}

Besides the background contributions listed above, the channels $\pi^+\pi^-\pi^0\gamma$ (after selection $<\SI{0.2}{\percent}$ compared to signal) and $\pi^+\pi^-4\pi^0\gamma$ (after selection $<\SI{0.1}{\percent}$ compared to signal) are subtracted using the generator AfkQed.

The generator KK2f~\cite{Jadach:1999vf} is used for the final state $\tau^+\tau^-$ but after the event selection less than 10 events remain to be subtracted, shown in Fig.~\ref{fig:allbkg}.
Other background contributions are negligible. 

\subsection{Alternative Method: Sideband Subtraction}\label{ssec:novobkg}
\begin{figure}
  \centering
  \includegraphics[type=pdf,ext=.pdf,read=.pdf, width=\linewidth]{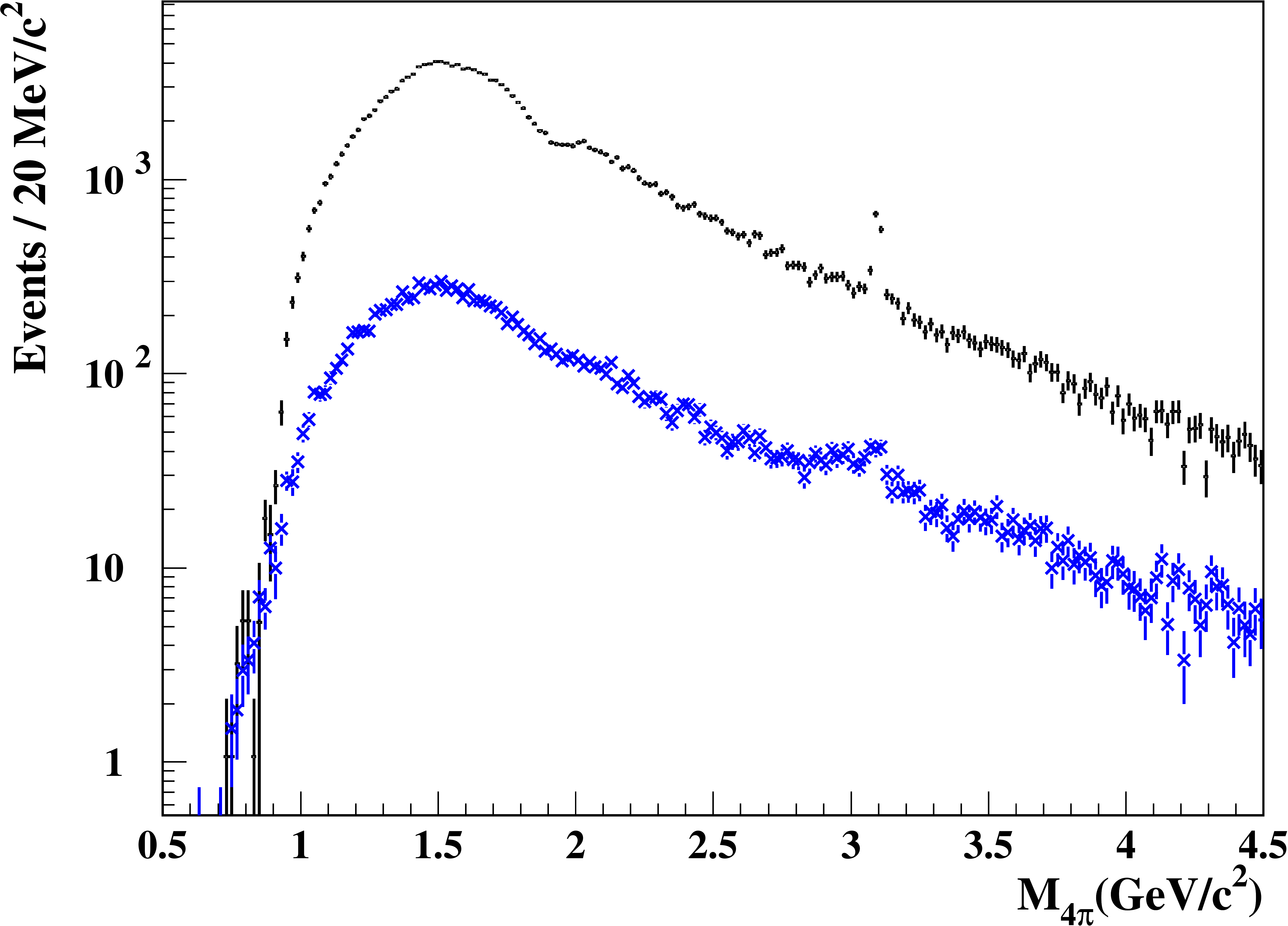}
  \colorcaption{Plot of the mass spectrum of the measured signal event rate (black points) before subtracting the background (blue crosses) using the sideband procedure.}
  \label{fig:sbbkg}
\end{figure}

The sideband subtraction method is a statistical procedure based on the $\chi^2_{\pi^+\pi^-2\pi^0\gamma}$ distribution of the kinematic fit to determine the appropriate number of events to subtract in each mass bin. The number of signal events is calculated as
\begin{equation}
  N_\mathrm{1s} = \frac{\beta}{\beta - \alpha} N_1 - \frac{1}{\beta - \alpha} N_2 \text{,}
\end{equation}
where $N_1$ and $N_2$ are the measured event numbers in the signal ($\chi^2 \le 30$) and sideband ($30 < \chi^2 < 60$) regions, respectively, such that $\alpha = N_{2\text{s}}/N_{1\text{s}}$ with events purely from the signal channel and $\beta = N_{2\text{b}}/N_{1\text{b}}$ with events purely from background. The signal $\chi^2$-distribution is taken from simulation, while the background is modeled by the difference between data and signal simulation (normalized at very low $\chi^2$), hence no background simulation is used. The background contribution from continuum processes is subtracted beforehand.
The resulting background level compared to data is shown in Fig.~\ref{fig:sbbkg} as a function of $M(\pi^+\pi^-2\pi^0)$.

\subsection{Comparison and Systematic Uncertainties}\label{ssec:bkgerr}

\begin{figure}
  \centering
  \includegraphics[type=pdf,ext=.pdf,read=.pdf, width=\linewidth]{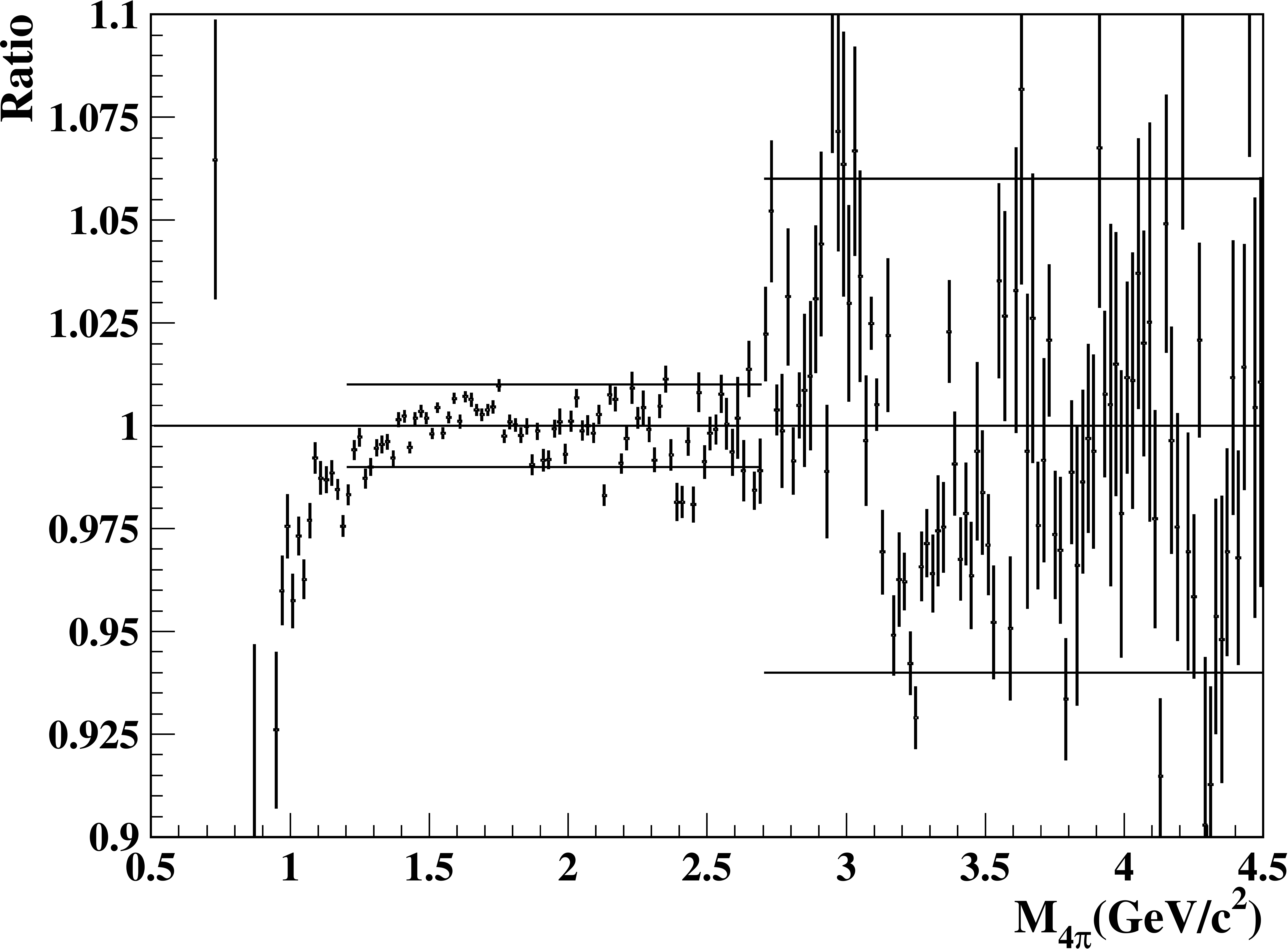}
  \caption{Ratio of the cross sections measured by adopting a background removal procedure based on simulation and on the sideband subtraction method. The horizontal lines indicate the systematic uncertainties for the subtraction of ISR backgrounds.}
  \label{fig:2pi2pi0csabreldiff}
\end{figure}

The two independent methods of subtracting the remaining background outlined above are compared in order to estimate the corresponding systematic uncertainty.
In the calculation of the $\pi^+\pi^-2\pi^0$ cross section the background subtraction procedure based on simulation is used.
The relative difference of the result from the sideband method is shown in Fig.~\ref{fig:2pi2pi0csabreldiff}. From this distribution, systematic uncertainties of \SI{1.0}{\percent} in the region $\threshul < M(\pi^+\pi^-2\pi^0) < \SI{2.7}{\GeVpercsq}$, and \SI{6.0}{\percent} for $M(\pi^+\pi^-2\pi^0) > \SI{2.7}{\GeVpercsq}$ are determined. For $\SI{0.85}{\GeVpercsq} \le M(\pi^+\pi^-2\pi^0) < \threshul$ the systematic uncertainty due to background subtraction is determined for each bin individually from the difference between the two subtraction methods.

\section{Acceptance and Efficiencies}\label{sec:eff}

\begin{figure}
  \centering
  \includegraphics[width=\linewidth]{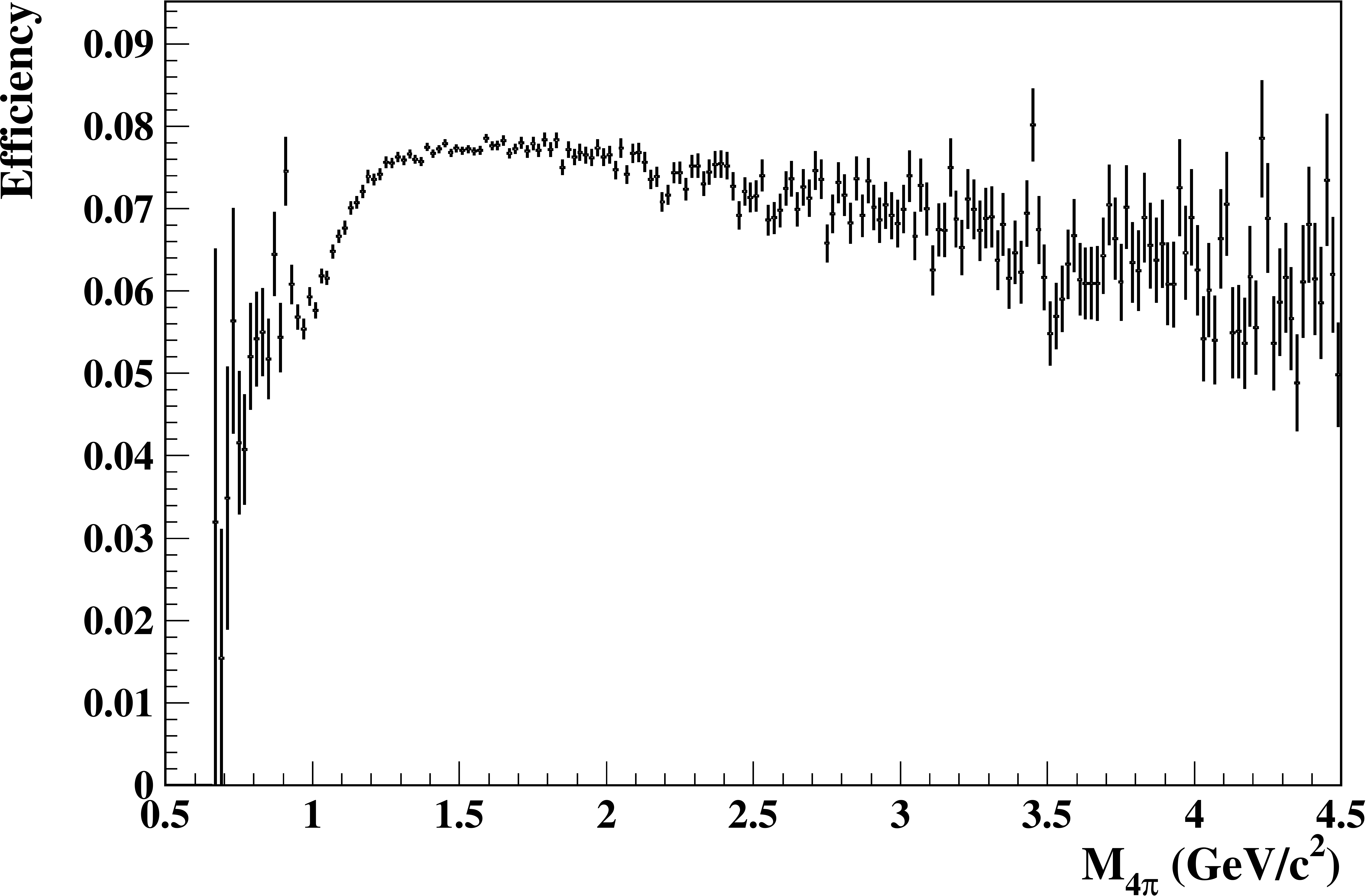}
  \caption{The simulated efficiency as a function of the $\pi^+\pi^-2\pi^0$ invariant mass.}
  \label{fig:acc}
\end{figure}

In order to calculate the efficiency of detecting a $\pi^+\pi^-2\pi^0\gamma$ event with the ISR photon generated in the angular range $| \cos{(\theta_\gamma^\ast)} | < C = \num{0.94}$ as a function of $M(\pi^+\pi^-2\pi^0)$, the detector simulation and event selection are applied to signal simulation. The result is subsequently divided by the number of events before selection, yielding the global efficiency shown in Fig.~\ref{fig:acc}. 
The sharp drop observed at low invariant masses is due to the kinematics of the ISR process. Low invariant masses correspond to a very high energetic ISR photon. Momentum conservation then dictates that the hadronic system must be emitted in a relatively small cone in the opposite direction of the ISR photon.
Therefore, at small hadronic invariant masses the inefficiency due to overlapping tracks or photons is increased.
Because ISR photons are radiated mostly at small polar angles, the probability of losing part of the hadronic system to the non-fiducial volume of the detector is significantly enhanced at small invariant masses.

\subsection{Photon efficiency}
In order to correct for inactive material, nonfunctioning crystals, and other sources of inefficiency in the photon detection, which may not be included in simulation, a detailed study is performed~\cite{Lees:2012cr}. For this purpose, the photon in $\mu^+\mu^-\gamma$ events is predicted based on the kinematic information from the charged tracks. The probability to detect the predicted photon is then compared between data and simulation. The result is used to correct the detection efficiency of every event as a function of the polar angle of the ISR photon. As a function of $M(\pi^+\pi^-2\pi^0)$, a uniform inefficiency difference of $\Delta\varepsilon_\gamma(\mathrm{MC} - \mathrm{data}) = \SI[parse-numbers=false]{(1.2 \pm 0.4)}{\percent}$ is observed and the total detection efficiency calculated in simulation is reduced accordingly.

\subsection{Tracking efficiency}

Efficiency differences between data and MC are also observed in track reconstruction. This is investigated using $e^+e^- \to \pi^+\pi^-\pi^+\pi^-\gamma$ events with one missing track~\cite{Lees:2012cr}. The missing track is predicted using a kinematic fit and the detection efficiency for the missing track is obtained in data and MC. Due to imperfect description of track overlap, small differences uniform in polar angle and transverse momentum exist. These yield a tracking efficiency correction of $\Delta \varepsilon_\mathrm{tr}(\mathrm{MC} - \mathrm{data}) = \SI[parse-numbers=false]{(0.9 \pm 0.8)}{\percent}$ for both tracks combined, slightly reducing the total detection efficiency calculated in simulation.

\subsection{\boldmath$\pi^0$ efficiency}

The probability of detecting a $\pi^0$ is studied extensively to uncover possible discrepancies between data and simulation which would need to be corrected.
In the ISR process $e^+e^- \to \omega\pi^0\gamma$, the unmeasured $\pi^0$ from the decay $\omega \to \pi^+\pi^-\pi^0$ can be inferred by a kinematic fit. The $\pi^0$ reconstruction efficiency is then determined as the fraction of events in the $\omega$ peak of the $M(\pi^+\pi^-\pi^0_\text{fit})$ distribution in which the $\pi^0$ has been detected.
This method is applied to data and simulation to determine differences between them. The resulting $\pi^0$ detection efficiencies yield an efficiency correction of $\Delta\varepsilon_{\pi^0}(\mathrm{MC} - \mathrm{data}) = \SI[parse-numbers=false]{(3.0 \pm 1.0)}{\percent}$ per $\pi^0$~\cite{Lees:2011zi}, which reduces the total detection efficiency calculated in simulation and has been studied to be valid in the full angular and momentum range.

\subsection{\boldmath$\chi^2_{\pi^+\pi^-2\pi^0\gamma}$ selection efficiency}

\begin{figure}
  \centering
  \includegraphics[width=\linewidth]{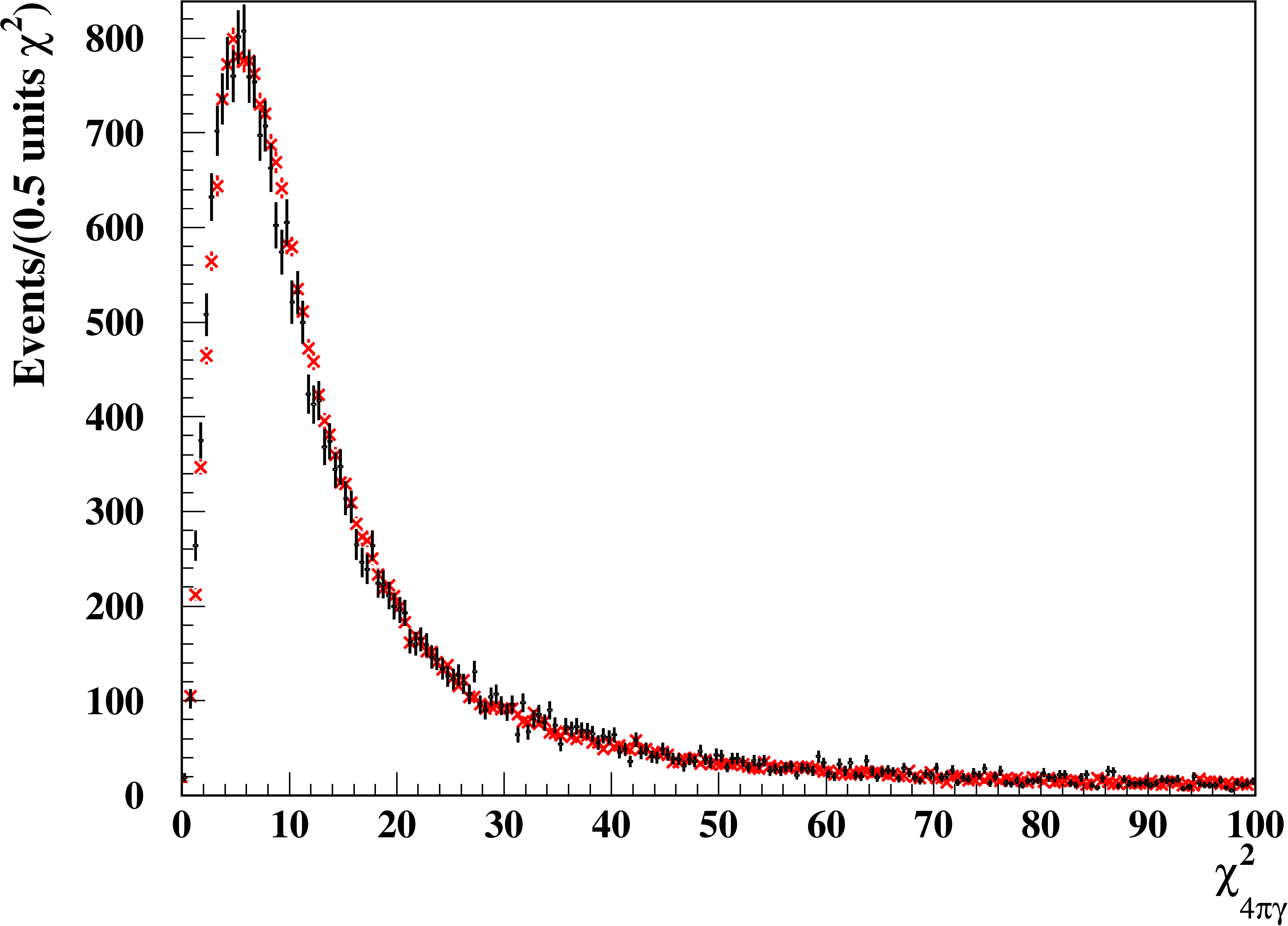}
  \caption{The $\chi^2_{\pi^+\pi^-2\pi^0\gamma}$ distributions in the clean sample for data (black points) and the AfkQed generator (red crosses, normalized to the same area as data in the range $\chi^2_{\pi^+\pi^-2\pi^0\gamma} < 10$).}
  \label{fig:chi2test}
\end{figure}

The choice of $\chi^2_{\pi^+\pi^-2\pi^0\gamma} < 30$ is studied by varying this requirement between \num{20} and \num{40}, yielding relative differences up to \SI{0.4}{\percent}, which is consequently used as the associated uncertainty. This uncertainty is confirmed in a study over a wider range up to $\chi^2_{\pi^+\pi^-2\pi^0\gamma} = 100$, which uses a clean event sample requiring exactly five photons in the final state in addition to the usual selection. The result is shown in Fig.~\ref{fig:chi2test}, where very good agreement between the $\chi^2_{\pi^+\pi^-2\pi^0\gamma}$ distributions in data and simulation is observed.

\section{Cross Section}\label{sec:cs}

\begin{figure}
  \centering
  \includegraphics[width=\linewidth]{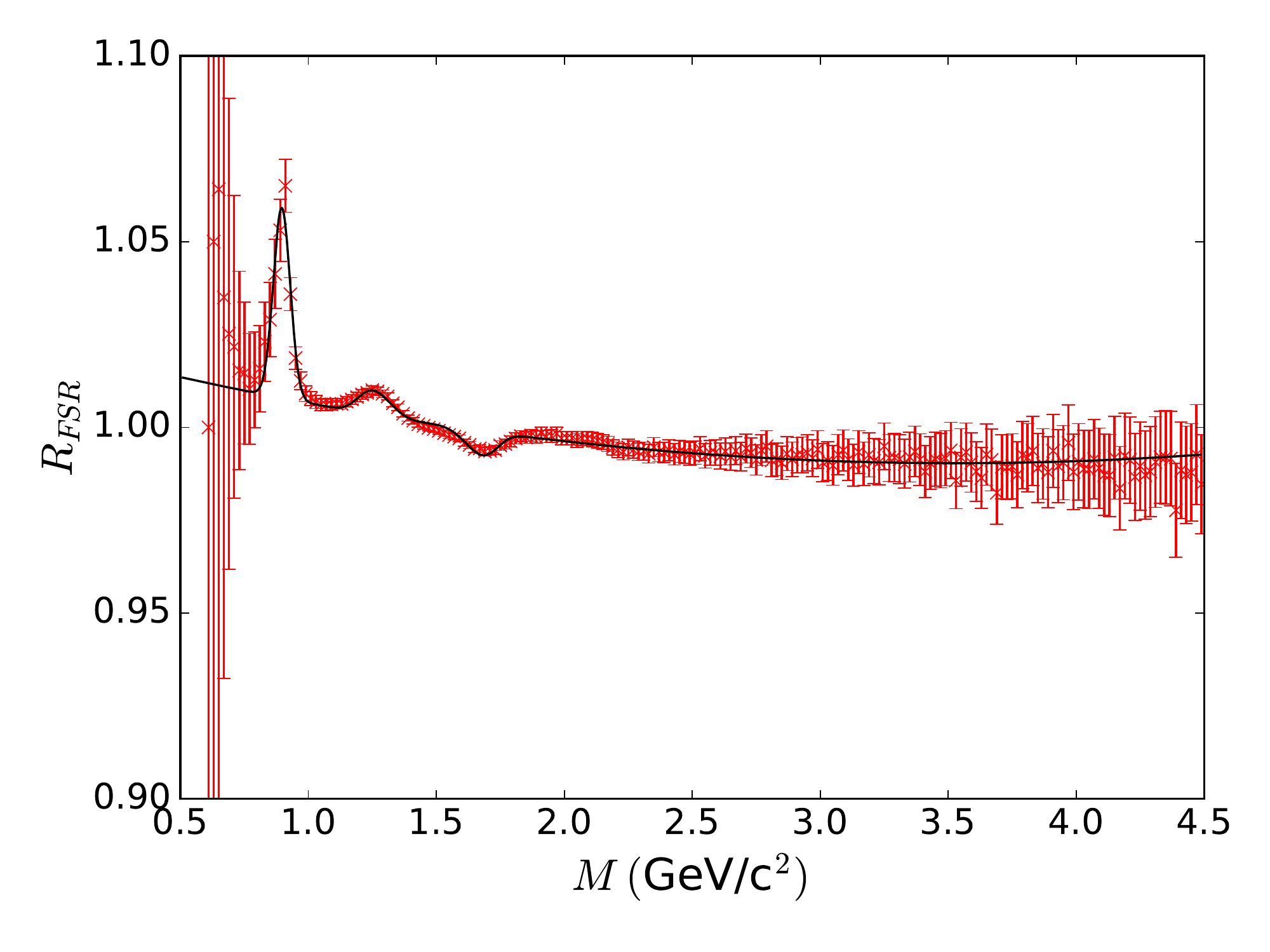}
  \colorcaption{Ratio of the simulated mass distributions including FSR+ISR and ISR only. The black line depicts a phenomenological fit function.}
  \label{fig:FSRcorr}
\end{figure}

The main purpose of this analysis is to determine the non-radiative cross section from the measured event rate:
\begin{equation}
  \sigma_{\pi^+\pi^-2\pi^0}(M) = \frac{\mathrm{d}N_{\pi^+\pi^-2\pi^0\gamma}(M)}{\mathrm{d}\mathcal{L}(M) \cdot \epsilon(M) (1+\delta(M))} \text{.}
  \label{eq:nonradcs}
\end{equation}
Here, $M \equiv M(\pi^+\pi^-2\pi^0)$, $\mathrm{d}N_{\pi^+\pi^-2\pi^0\gamma}$ is the number of events after selection and background subtraction in the interval $\mathrm{d}M$, $\mathrm{d}\mathcal{L}$ the differential ISR-luminosity, $\epsilon(M)$ the combined acceptance and efficiency, and $\delta$ the correction for radiative effects including FSR.
The AfkQed generator used in combination with the detector simulation contains corrections for NLO-ISR collinear to the beam as well as FSR corrections implemented by PHOTOS~\cite{Barberio:1993qi}. 
The NLO-ISR correction is calculated by comparing the generator with PHOKHARA~\cite{Czyz:2008kw}, which includes the full ISR contributions up to NLO. An effect of \SI[parse-numbers=false]{(0.8 \pm 0.1_\mathrm{stat} \pm 0.5_\mathrm{syst})}{\percent}, constant in $M(\pi^+\pi^-2\pi^0)$, is observed and subsequently corrected for.
Final-state radiation shifts events towards smaller invariant masses. Therefore, a mass-dependent correction is applied corresponding to the relative change in the content of each mass bin. This is calculated by dividing the simulated event rate with FSR by the event rate without FSR, as shown in Fig.~\ref{fig:FSRcorr}. The measured event distribution is then divided by the phenomenological fit function to reverse the effect of FSR.
Besides radiative effects, the mass resolution is considered in the cross section measurement. The invariant mass $M(\pi^+\pi^-2\pi^0)$ has a resolution of \SI{15}{\MeVpercsq} in the range of interest. 
Since the cross section is given in bins of \SI{20}{\MeVpercsq}, events with nominal bin-center mass are distributed such that \SI{50}{\percent} will lie in the central bin, \SI{23}{\percent} in each neighboring bin, and \SI{2}{\percent} in the next bins.
The effect of the mass resolution has been studied by performing unfolding procedures based on singular value decomposition~\cite{Hocker:1995kb} and Tikhonov regularized $\chi^2$ minimization with L-curve optimization~\cite{Schmitt:2012kp}. It is observed that the effect of the mass resolution is consistent with zero with a systematic uncertainty of \SI{0.3}{\percent}.

\begin{figure}
  \centering
  \includegraphics[type=pdf,ext=.pdf,read=.pdf, width=\linewidth]{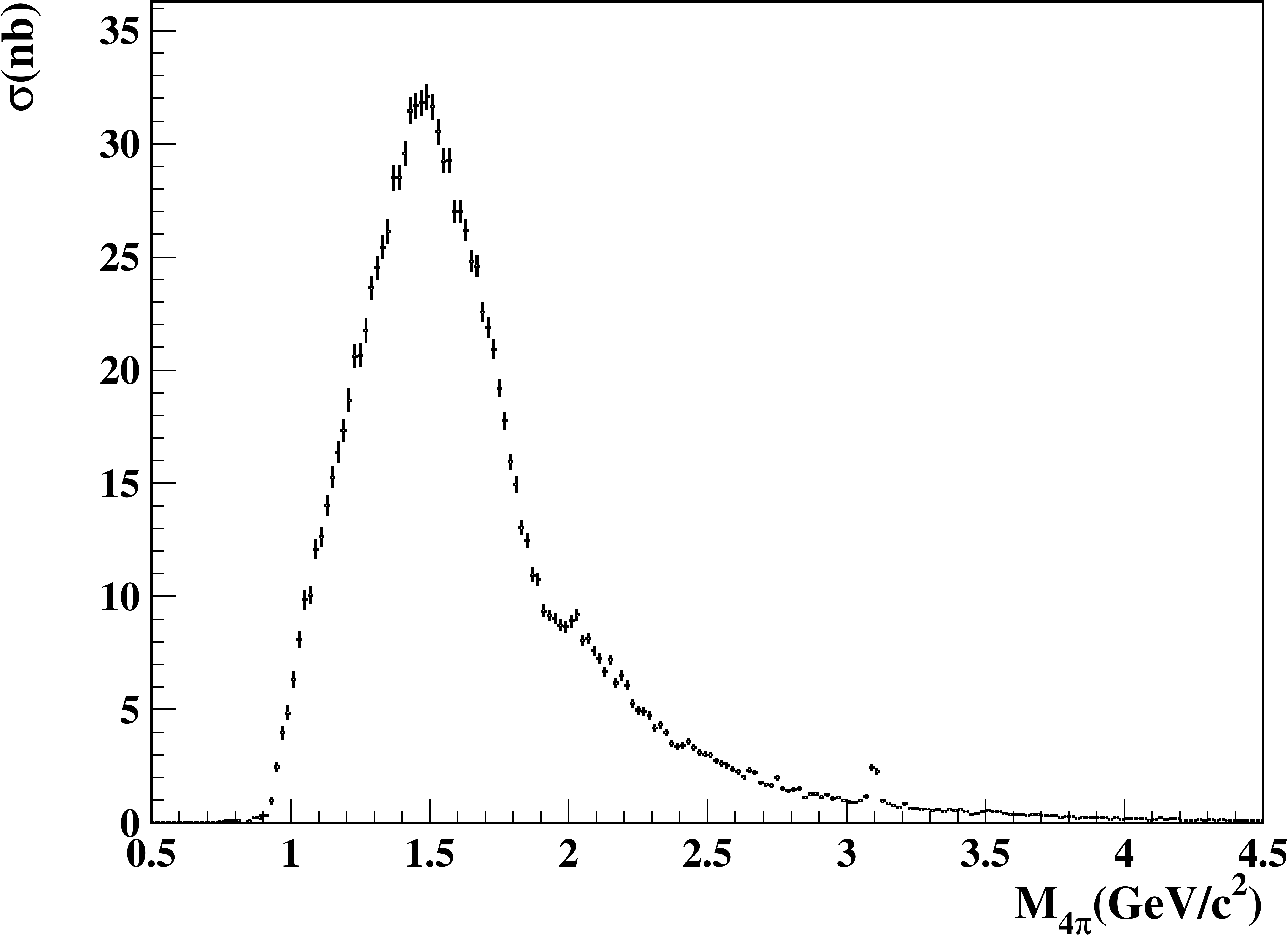}
  \caption{The measured dressed $\pi^+\pi^-2\pi^0$ cross section (statistical uncertainties only).}
  \label{fig:cs}
\end{figure}

Once all corrections are applied and the efficiency is determined (including data-MC differences from photon, track and $\pi^0$ detection), Eq.~(\ref{eq:nonradcs}) is employed to calculate the non-radiative cross section $\sigma$, displayed in Fig.~\ref{fig:cs} and listed in Table~\ref{tab:csall}.

Removing the effect of vacuum polarization (VP) leads to the \emph{undressed} cross section $\sigma^\mathrm{(0)}$, which is related to its originally \emph{dressed} equivalent $\sigma$ through the transformation~\cite{Eidelman:1995ny}
\begin{equation}
  \sigma^\mathrm{(0)}_{\pi^+\pi^-2\pi^0}(E_\mathrm{CM}) = \sigma_{\pi^+\pi^-2\pi^0}(E_\mathrm{CM}) \cdot \left( \frac{\alpha(0)}{\alpha(E_\mathrm{CM})} \right)^2 \text{,}
  \label{eq:undressedcs}
\end{equation}
where $\alpha$ is the QED coupling at the center-of-mass energy $E_\mathrm{CM}$, with $\alpha(0) = \num[separate-uncertainty=false]{7.2973525664(17)e-3}$~\cite{PDG}. The undressed cross section is also listed in Table~\ref{tab:csall}.

\subsection{Systematic Uncertainties}

Table~\ref{tab:syst} shows the systematic uncertainties in this analysis. 

The efficiency predicted by the Monte Carlo generator AfkQed is affected by the relative weight of the resonances included in the simulation. The model used in AfkQed includes the $\rho$, $\rho^\prime$, and $\rho^{\prime\prime}$ resonances as well as the intermediate states $\omega\pi^0$, $a_1(1260)\pi$, and a small contribution from $\rho^0f_0$. The corresponding uncertainty due to their relative weight was determined to be less than \SI{0.4}{\percent}.

The normalization of the continuum simulation introduces an uncertainty which translates to \SI{2.0}{\percent} in the mass range above \SI{3.2}{\GeVpercsq} and \SI{1.0}{\percent} below.
The PID algorithms in this analysis generate \SI{0.5}{\percent} uncertainty from the kaon identification and \SI{2.0}{\percent} uncertainty from the combined muon veto above \SI{3.2}{\GeVpercsq}.

Assuming these effects to be uncorrelated, the total systematic uncertainties listed in Table~\ref{tab:syst} are found in different mass regions.
For $M(\pi^+\pi^-2\pi^0) \leq \threshul$, the systematic uncertainty due to ISR background subtraction is determined bin by bin and ranges from \SI{1}{\percent} to \SI{100}{\percent}.
In this region the absolute systematic uncertainty due to ISR background subtraction is calculated as $\SI[parse-numbers=false]{(0.455 \cdot E_\text{CM}/\si{GeV} - 0.296)}{\nano\barn}$.
In the region below \SI{0.85}{\GeVpercsq} the measurement is compatible with zero.

  \afterpage{
    \LTcapwidth=\linewidth
\begin{longtable}{c r r}
\caption{The measured $e^+e^- \to \pi^+\pi^-\pi^0\pi^0$ cross section. The \textit{dressed} $\sigma$ (including VP) and the \textit{undressed} $\sigma^{(0)}$ (without VP) cross sections are reported separately, each with the corresponding statistical uncertainties.} \\
\multicolumn{1}{c}{$E_\mathrm{CM} (\si{\GeV})$} & \multicolumn{1}{c}{$\sigma (\si{\nano\barn})$} & \multicolumn{1}{c}{$\sigma^{(0)} (\si{\nano\barn})$} \\
\hline
\endfirsthead
\multicolumn{1}{c}{$E_\mathrm{CM} (\si{\GeV})$} & \multicolumn{1}{c}{$\sigma (\si{\nano\barn})$} & \multicolumn{1}{c}{$\sigma^{(0)} (\si{\nano\barn})$} \\
\hline
\endhead
0.85 & 0.05 $\pm$ 0.12 & 0.05 $\pm$ 0.11 \\
0.87 & 0.24 $\pm$ 0.08 & 0.23 $\pm$ 0.07 \\
0.89 & 0.23 $\pm$ 0.12 & 0.22 $\pm$ 0.12 \\
0.91 & 0.31 $\pm$ 0.08 & 0.30 $\pm$ 0.07 \\
0.93 & 0.98 $\pm$ 0.16 & 0.95 $\pm$ 0.16 \\
0.95 & 2.46 $\pm$ 0.23 & 2.38 $\pm$ 0.23 \\
0.97 & 3.98 $\pm$ 0.31 & 3.86 $\pm$ 0.30 \\
0.99 & 4.86 $\pm$ 0.32 & 4.75 $\pm$ 0.32 \\
1.01 & 6.32 $\pm$ 0.38 & 6.41 $\pm$ 0.39 \\
1.03 & 8.09 $\pm$ 0.40 & 7.43 $\pm$ 0.37 \\
1.05 & 9.85 $\pm$ 0.42 & 9.32 $\pm$ 0.40 \\
1.07 & 10.06 $\pm$ 0.42 & 9.59 $\pm$ 0.40 \\
1.09 & 12.08 $\pm$ 0.44 & 11.56 $\pm$ 0.43 \\
1.11 & 12.62 $\pm$ 0.45 & 12.10 $\pm$ 0.43 \\
1.13 & 14.02 $\pm$ 0.47 & 13.47 $\pm$ 0.45 \\
1.15 & 15.26 $\pm$ 0.48 & 14.67 $\pm$ 0.47 \\
1.17 & 16.39 $\pm$ 0.48 & 15.77 $\pm$ 0.47 \\
1.19 & 17.33 $\pm$ 0.49 & 16.69 $\pm$ 0.47 \\
1.21 & 18.66 $\pm$ 0.53 & 17.98 $\pm$ 0.51 \\
1.23 & 20.62 $\pm$ 0.52 & 19.89 $\pm$ 0.51 \\
1.25 & 20.66 $\pm$ 0.52 & 19.93 $\pm$ 0.50 \\
1.27 & 21.75 $\pm$ 0.55 & 21.00 $\pm$ 0.53 \\
1.29 & 23.62 $\pm$ 0.54 & 22.81 $\pm$ 0.52 \\
1.31 & 24.51 $\pm$ 0.55 & 23.68 $\pm$ 0.53 \\
1.33 & 25.43 $\pm$ 0.55 & 24.57 $\pm$ 0.53 \\
1.35 & 26.13 $\pm$ 0.56 & 25.25 $\pm$ 0.54 \\
1.37 & 28.49 $\pm$ 0.58 & 27.54 $\pm$ 0.56 \\
1.39 & 28.50 $\pm$ 0.57 & 27.55 $\pm$ 0.55 \\
1.41 & 29.56 $\pm$ 0.57 & 28.58 $\pm$ 0.55 \\
1.43 & 31.45 $\pm$ 0.59 & 30.41 $\pm$ 0.57 \\
1.45 & 31.66 $\pm$ 0.59 & 30.62 $\pm$ 0.57 \\
1.47 & 31.80 $\pm$ 0.59 & 30.75 $\pm$ 0.57 \\
1.49 & 32.07 $\pm$ 0.58 & 31.00 $\pm$ 0.56 \\
1.51 & 31.64 $\pm$ 0.57 & 30.57 $\pm$ 0.56 \\
1.53 & 30.53 $\pm$ 0.56 & 29.48 $\pm$ 0.54 \\
1.55 & 29.24 $\pm$ 0.55 & 28.21 $\pm$ 0.53 \\
1.57 & 29.26 $\pm$ 0.55 & 28.23 $\pm$ 0.53 \\
1.59 & 27.01 $\pm$ 0.51 & 26.05 $\pm$ 0.49 \\
1.61 & 27.02 $\pm$ 0.51 & 26.06 $\pm$ 0.49 \\
1.63 & 26.19 $\pm$ 0.50 & 25.26 $\pm$ 0.48 \\
1.65 & 24.80 $\pm$ 0.48 & 23.91 $\pm$ 0.46 \\
1.67 & 24.60 $\pm$ 0.48 & 23.71 $\pm$ 0.46 \\
1.69 & 22.56 $\pm$ 0.46 & 21.73 $\pm$ 0.44 \\
1.71 & 21.89 $\pm$ 0.45 & 21.07 $\pm$ 0.43 \\
1.73 & 20.93 $\pm$ 0.44 & 20.14 $\pm$ 0.43 \\
1.75 & 19.20 $\pm$ 0.42 & 18.47 $\pm$ 0.40 \\
1.77 & 17.76 $\pm$ 0.41 & 17.08 $\pm$ 0.39 \\
1.79 & 15.94 $\pm$ 0.38 & 15.33 $\pm$ 0.36 \\
1.81 & 14.94 $\pm$ 0.37 & 14.37 $\pm$ 0.35 \\
1.83 & 13.03 $\pm$ 0.34 & 12.53 $\pm$ 0.32 \\
1.85 & 12.47 $\pm$ 0.34 & 11.99 $\pm$ 0.32 \\
1.87 & 10.95 $\pm$ 0.31 & 10.53 $\pm$ 0.29 \\
1.89 & 10.74 $\pm$ 0.30 & 10.34 $\pm$ 0.29 \\
1.91 & 9.36 $\pm$ 0.28 & 9.02 $\pm$ 0.27 \\
1.93 & 9.15 $\pm$ 0.28 & 8.81 $\pm$ 0.26 \\
1.95 & 9.01 $\pm$ 0.27 & 8.68 $\pm$ 0.26 \\
1.97 & 8.72 $\pm$ 0.26 & 8.40 $\pm$ 0.26 \\
1.99 & 8.66 $\pm$ 0.26 & 8.35 $\pm$ 0.26 \\
2.01 & 8.91 $\pm$ 0.27 & 8.59 $\pm$ 0.26 \\
2.03 & 9.17 $\pm$ 0.28 & 8.84 $\pm$ 0.27 \\
2.05 & 8.06 $\pm$ 0.25 & 7.76 $\pm$ 0.24 \\
2.07 & 8.14 $\pm$ 0.26 & 7.84 $\pm$ 0.25 \\
2.09 & 7.59 $\pm$ 0.24 & 7.31 $\pm$ 0.23 \\
2.11 & 7.27 $\pm$ 0.23 & 7.00 $\pm$ 0.23 \\
2.13 & 6.67 $\pm$ 0.23 & 6.42 $\pm$ 0.22 \\
2.15 & 7.20 $\pm$ 0.24 & 6.92 $\pm$ 0.23 \\
2.17 & 6.17 $\pm$ 0.22 & 5.93 $\pm$ 0.21 \\
2.19 & 6.51 $\pm$ 0.23 & 6.26 $\pm$ 0.22 \\
2.21 & 6.09 $\pm$ 0.22 & 5.85 $\pm$ 0.21 \\
2.23 & 5.28 $\pm$ 0.20 & 5.08 $\pm$ 0.19 \\
2.25 & 4.96 $\pm$ 0.19 & 4.77 $\pm$ 0.18 \\
2.27 & 4.91 $\pm$ 0.19 & 4.72 $\pm$ 0.19 \\
2.29 & 4.75 $\pm$ 0.18 & 4.56 $\pm$ 0.18 \\
2.31 & 4.19 $\pm$ 0.17 & 4.03 $\pm$ 0.16 \\
2.33 & 4.34 $\pm$ 0.18 & 4.16 $\pm$ 0.17 \\
2.35 & 3.98 $\pm$ 0.17 & 3.82 $\pm$ 0.16 \\
2.37 & 3.50 $\pm$ 0.15 & 3.36 $\pm$ 0.15 \\
2.39 & 3.37 $\pm$ 0.15 & 3.24 $\pm$ 0.15 \\
2.41 & 3.42 $\pm$ 0.15 & 3.28 $\pm$ 0.15 \\
2.43 & 3.59 $\pm$ 0.16 & 3.45 $\pm$ 0.15 \\
2.45 & 3.33 $\pm$ 0.16 & 3.20 $\pm$ 0.15 \\
2.47 & 3.09 $\pm$ 0.15 & 2.97 $\pm$ 0.14 \\
2.49 & 3.02 $\pm$ 0.15 & 2.90 $\pm$ 0.14 \\
2.51 & 2.99 $\pm$ 0.15 & 2.87 $\pm$ 0.14 \\
2.53 & 2.72 $\pm$ 0.14 & 2.62 $\pm$ 0.13 \\
2.55 & 2.61 $\pm$ 0.14 & 2.51 $\pm$ 0.13 \\
2.57 & 2.53 $\pm$ 0.14 & 2.44 $\pm$ 0.13 \\
2.59 & 2.36 $\pm$ 0.13 & 2.27 $\pm$ 0.12 \\
2.61 & 2.26 $\pm$ 0.13 & 2.17 $\pm$ 0.12 \\
2.63 & 2.01 $\pm$ 0.12 & 1.94 $\pm$ 0.11 \\
2.65 & 2.34 $\pm$ 0.13 & 2.25 $\pm$ 0.12 \\
2.67 & 2.22 $\pm$ 0.12 & 2.13 $\pm$ 0.12 \\
2.69 & 1.76 $\pm$ 0.11 & 1.70 $\pm$ 0.10 \\
2.71 & 1.67 $\pm$ 0.10 & 1.61 $\pm$ 0.10 \\
2.73 & 1.65 $\pm$ 0.11 & 1.58 $\pm$ 0.10 \\
2.75 & 2.00 $\pm$ 0.12 & 1.92 $\pm$ 0.12 \\
2.77 & 1.49 $\pm$ 0.10 & 1.44 $\pm$ 0.10 \\
2.79 & 1.39 $\pm$ 0.10 & 1.34 $\pm$ 0.09 \\
2.81 & 1.46 $\pm$ 0.10 & 1.41 $\pm$ 0.09 \\
2.83 & 1.50 $\pm$ 0.10 & 1.44 $\pm$ 0.10 \\
2.85 & 1.10 $\pm$ 0.08 & 1.07 $\pm$ 0.08 \\
2.87 & 1.26 $\pm$ 0.09 & 1.22 $\pm$ 0.09 \\
2.89 & 1.27 $\pm$ 0.09 & 1.23 $\pm$ 0.09 \\
2.91 & 1.15 $\pm$ 0.09 & 1.11 $\pm$ 0.09 \\
2.93 & 1.22 $\pm$ 0.09 & 1.19 $\pm$ 0.09 \\
2.95 & 1.07 $\pm$ 0.09 & 1.04 $\pm$ 0.09 \\
2.97 & 1.12 $\pm$ 0.09 & 1.09 $\pm$ 0.09 \\
2.99 & 1.00 $\pm$ 0.09 & 0.97 $\pm$ 0.08 \\
3.01 & 0.91 $\pm$ 0.08 & 0.90 $\pm$ 0.08 \\
3.03 & 0.91 $\pm$ 0.08 & 0.90 $\pm$ 0.08 \\
3.05 & 0.98 $\pm$ 0.08 & 0.99 $\pm$ 0.08 \\
3.07 & 1.17 $\pm$ 0.09 & 1.21 $\pm$ 0.09 \\
3.09 & 2.44 $\pm$ 0.15 & 3.14 $\pm$ 0.19 \\
3.11 & 2.28 $\pm$ 0.15 & 1.83 $\pm$ 0.12 \\
3.13 & 0.95 $\pm$ 0.08 & 0.85 $\pm$ 0.07 \\
3.15 & 0.87 $\pm$ 0.08 & 0.80 $\pm$ 0.07 \\
3.17 & 0.76 $\pm$ 0.06 & 0.71 $\pm$ 0.06 \\
3.19 & 0.69 $\pm$ 0.06 & 0.64 $\pm$ 0.06 \\
3.21 & 0.82 $\pm$ 0.07 & 0.77 $\pm$ 0.07 \\
3.23 & 0.65 $\pm$ 0.06 & 0.61 $\pm$ 0.06 \\
3.25 & 0.65 $\pm$ 0.06 & 0.61 $\pm$ 0.06 \\
3.27 & 0.59 $\pm$ 0.06 & 0.55 $\pm$ 0.05 \\
3.29 & 0.62 $\pm$ 0.06 & 0.59 $\pm$ 0.06 \\
3.31 & 0.54 $\pm$ 0.05 & 0.51 $\pm$ 0.05 \\
3.33 & 0.59 $\pm$ 0.06 & 0.56 $\pm$ 0.06 \\
3.35 & 0.47 $\pm$ 0.05 & 0.45 $\pm$ 0.05 \\
3.37 & 0.59 $\pm$ 0.06 & 0.56 $\pm$ 0.06 \\
3.39 & 0.54 $\pm$ 0.06 & 0.51 $\pm$ 0.05 \\
3.41 & 0.59 $\pm$ 0.06 & 0.56 $\pm$ 0.06 \\
3.43 & 0.47 $\pm$ 0.05 & 0.45 $\pm$ 0.05 \\
3.45 & 0.39 $\pm$ 0.04 & 0.37 $\pm$ 0.04 \\
3.47 & 0.41 $\pm$ 0.05 & 0.39 $\pm$ 0.05 \\
3.49 & 0.51 $\pm$ 0.06 & 0.48 $\pm$ 0.05 \\
3.51 & 0.55 $\pm$ 0.06 & 0.53 $\pm$ 0.06 \\
3.53 & 0.52 $\pm$ 0.06 & 0.50 $\pm$ 0.06 \\
3.55 & 0.46 $\pm$ 0.06 & 0.44 $\pm$ 0.05 \\
3.57 & 0.42 $\pm$ 0.05 & 0.40 $\pm$ 0.05 \\
3.59 & 0.37 $\pm$ 0.04 & 0.35 $\pm$ 0.04 \\
3.61 & 0.37 $\pm$ 0.05 & 0.35 $\pm$ 0.05 \\
3.63 & 0.38 $\pm$ 0.05 & 0.37 $\pm$ 0.05 \\
3.65 & 0.31 $\pm$ 0.04 & 0.31 $\pm$ 0.04 \\
3.67 & 0.35 $\pm$ 0.05 & 0.35 $\pm$ 0.05 \\
3.69 & 0.36 $\pm$ 0.04 & 0.27 $\pm$ 0.03 \\
3.71 & 0.31 $\pm$ 0.04 & 0.28 $\pm$ 0.04 \\
3.73 & 0.29 $\pm$ 0.04 & 0.27 $\pm$ 0.04 \\
3.75 & 0.32 $\pm$ 0.04 & 0.30 $\pm$ 0.04 \\
3.77 & 0.22 $\pm$ 0.03 & 0.20 $\pm$ 0.03 \\
3.79 & 0.28 $\pm$ 0.04 & 0.26 $\pm$ 0.04 \\
3.81 & 0.27 $\pm$ 0.04 & 0.25 $\pm$ 0.04 \\
3.83 & 0.18 $\pm$ 0.03 & 0.17 $\pm$ 0.03 \\
3.85 & 0.23 $\pm$ 0.03 & 0.22 $\pm$ 0.03 \\
3.87 & 0.25 $\pm$ 0.04 & 0.24 $\pm$ 0.04 \\
3.89 & 0.21 $\pm$ 0.03 & 0.20 $\pm$ 0.03 \\
3.91 & 0.21 $\pm$ 0.03 & 0.20 $\pm$ 0.03 \\
3.93 & 0.25 $\pm$ 0.04 & 0.24 $\pm$ 0.04 \\
3.95 & 0.14 $\pm$ 0.03 & 0.14 $\pm$ 0.02 \\
3.97 & 0.20 $\pm$ 0.03 & 0.19 $\pm$ 0.03 \\
3.99 & 0.14 $\pm$ 0.03 & 0.13 $\pm$ 0.02 \\
4.01 & 0.19 $\pm$ 0.03 & 0.18 $\pm$ 0.03 \\
4.03 & 0.18 $\pm$ 0.03 & 0.17 $\pm$ 0.03 \\
4.05 & 0.16 $\pm$ 0.03 & 0.16 $\pm$ 0.03 \\
4.07 & 0.18 $\pm$ 0.03 & 0.17 $\pm$ 0.03 \\
4.09 & 0.11 $\pm$ 0.02 & 0.10 $\pm$ 0.02 \\
4.11 & 0.15 $\pm$ 0.02 & 0.14 $\pm$ 0.02 \\
4.13 & 0.19 $\pm$ 0.03 & 0.18 $\pm$ 0.03 \\
4.15 & 0.16 $\pm$ 0.03 & 0.15 $\pm$ 0.03 \\
4.17 & 0.19 $\pm$ 0.03 & 0.18 $\pm$ 0.03 \\
4.19 & 0.16 $\pm$ 0.03 & 0.15 $\pm$ 0.03 \\
4.21 & 0.09 $\pm$ 0.02 & 0.08 $\pm$ 0.02 \\
4.23 & 0.10 $\pm$ 0.02 & 0.10 $\pm$ 0.02 \\
4.25 & 0.12 $\pm$ 0.02 & 0.11 $\pm$ 0.02 \\
4.27 & 0.16 $\pm$ 0.03 & 0.15 $\pm$ 0.03 \\
4.29 & 0.07 $\pm$ 0.02 & 0.07 $\pm$ 0.02 \\
4.31 & 0.13 $\pm$ 0.02 & 0.12 $\pm$ 0.02 \\
4.33 & 0.12 $\pm$ 0.03 & 0.12 $\pm$ 0.02 \\
4.35 & 0.13 $\pm$ 0.03 & 0.13 $\pm$ 0.03 \\
4.37 & 0.11 $\pm$ 0.02 & 0.11 $\pm$ 0.02 \\
4.39 & 0.08 $\pm$ 0.02 & 0.08 $\pm$ 0.02 \\
4.41 & 0.11 $\pm$ 0.02 & 0.10 $\pm$ 0.02 \\
4.43 & 0.12 $\pm$ 0.02 & 0.11 $\pm$ 0.02 \\
4.45 & 0.07 $\pm$ 0.02 & 0.07 $\pm$ 0.02 \\
4.47 & 0.08 $\pm$ 0.02 & 0.07 $\pm$ 0.02 \\
4.49 & 0.09 $\pm$ 0.02 & 0.08 $\pm$ 0.02 \\
\hline
\vspace{-3ex}
\label{tab:csall}
\end{longtable}
    }
  \afterpage{
  \begin{table}
    \centering
    \caption{Systematic uncertainties for different mass ranges. For the ISR background subtraction uncertainty in the low-mass region $M(\pi^+\pi^-2\pi^0) < \threshul$ see text.}
    \begin{tabular}{l r r r r}
      $M(\pi^+\pi^-2\pi^0) \si{(\GeVpercsq)}$ & $<1.2$ & $1.2$ -- $2.7$ & $2.7$ -- $3.2$ & ${} > 3.2$ \\
      \hline
      Tracking eff. & $\SI{0.8}{\percent}$ & $\SI{0.8}{\percent}$ & $\SI{0.8}{\percent}$ & $\SI{0.8}{\percent}$ \\
      $\gamma$ eff. & $\SI{0.4}{\percent}$ & $\SI{0.4}{\percent}$ & $\SI{0.4}{\percent}$ & $\SI{0.4}{\percent}$ \\
      $2\pi^0$ eff. & $\SI{2.0}{\percent}$ & $\SI{2.0}{\percent}$ & $\SI{2.0}{\percent}$ & $\SI{2.0}{\percent}$ \\
      $\chi^2_{\pi^+\pi^-2\pi^0\gamma}$ eff. & $\SI{0.4}{\percent}$ & $\SI{0.4}{\percent}$ & $\SI{0.4}{\percent}$ & $\SI{0.4}{\percent}$ \\
      Generator model & $\SI{0.4}{\percent}$ & $\SI{0.4}{\percent}$ & $\SI{0.4}{\percent}$ & $\SI{0.4}{\percent}$ \\
      Mass res. & $\SI{0.3}{\percent}$ & $\SI{0.3}{\percent}$ & $\SI{0.3}{\percent}$ & $\SI{0.3}{\percent}$ \\
      \hline
      FSR & $\SI{1.0}{\percent}$ & $\SI{1.0}{\percent}$ & $\SI{1.0}{\percent}$ & $\SI{1.0}{\percent}$ \\
      NLO ISR & $\SI{0.5}{\percent}$ & $\SI{0.5}{\percent}$  & $\SI{0.5}{\percent}$ & $\SI{0.5}{\percent}$ \\
      ISR luminosity & $\SI{1.0}{\percent}$ & $\SI{1.0}{\percent}$ & $\SI{1.0}{\percent}$ & $\SI{1.0}{\percent}$ \\
      \hline
      Continuum Bkg & $\SI{1.0}{\percent}$ & $\SI{1.0}{\percent}$ & $\SI{1.0}{\percent}$ & $\SI{2.0}{\percent}$ \\
      ISR Background & $\SI[parse-numbers=false]{1 - 100}{\percent}$ & $\SI{1.0}{\percent}$ & $\SI{6.0}{\percent}$ & $\SI{6.0}{\percent}$ \\
      \hline
      Kaon PID & $\SI{0.5}{\percent}$ & $\SI{0.5}{\percent}$ & $\SI{0.5}{\percent}$ & $\SI{0.5}{\percent}$ \\
      Muon PID & $\SI{0}{\percent}$ &$\SI{0}{\percent}$ & $\SI{0}{\percent}$ & $\SI{2.0}{\percent}$ \\
      \hline
      total & $\SI[parse-numbers=false]{3 - 100}{\percent}$ & $\SI{3.1}{\percent}$ & $\SI{6.7}{\percent}$ & $\SI{7.2}{\percent}$
    \end{tabular}
    \label{tab:syst}
  \end{table}
  }
  
\subsection{Comparison to theory and other experiments}

\begin{figure}
  \centering
  \includegraphics[width=\linewidth]{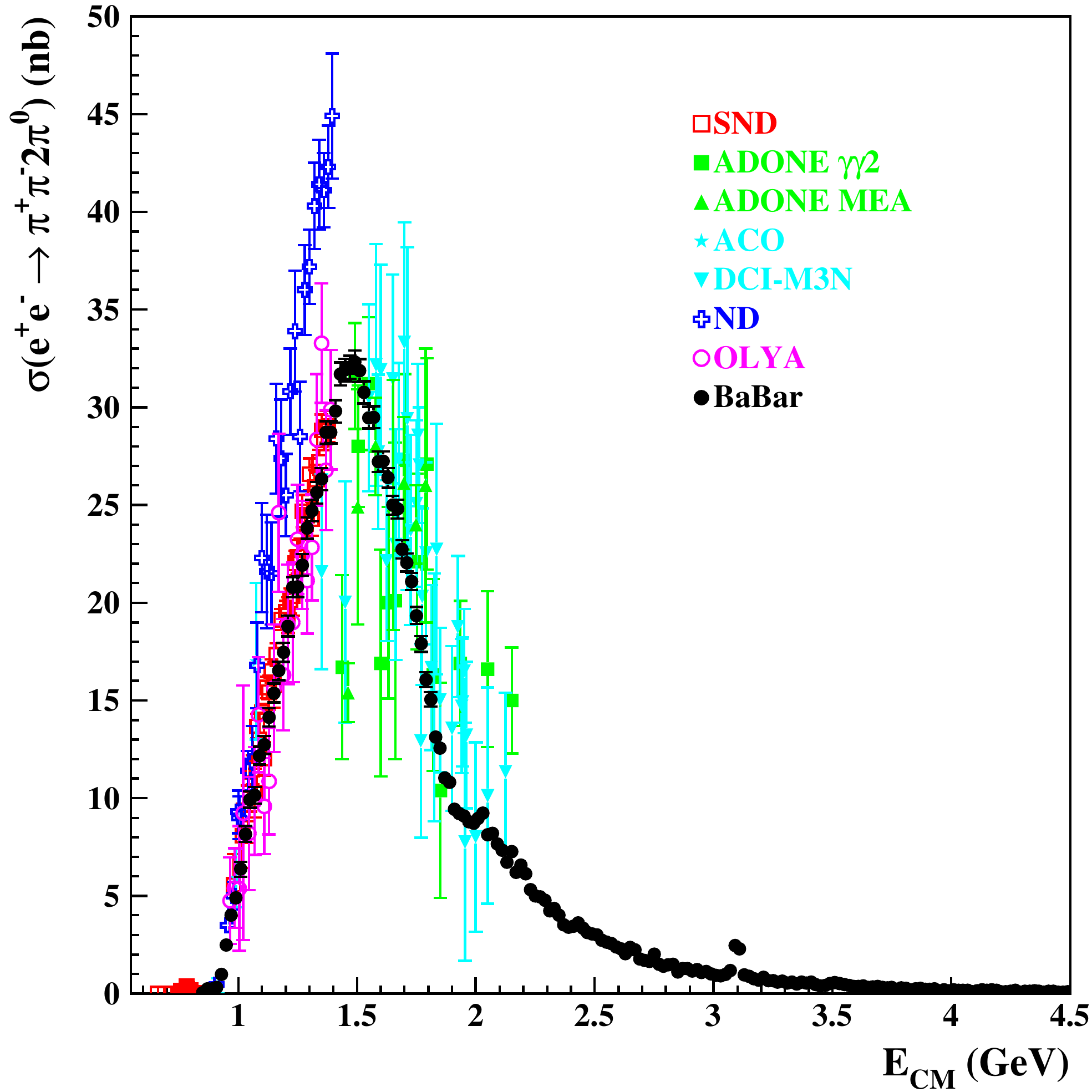}
  \colorcaption{The previously published $\pi^+\pi^-2\pi^0$ cross section data in addition to this analysis (statistical uncertainties only).}
  \label{fig:2pi2pi0compilation}
\end{figure}

The measured cross section is compared to existing data in Fig.~\ref{fig:2pi2pi0compilation}. Our new measurement covers the energy range from \SI{0.85}{\GeV} to \SI{4.5}{\GeV}. The previously existing data was collected by the experiments ACO~\cite{Cosme:1972wt,Cosme:1976tf}, ADONE MEA~\cite{Esposito:1977ct,Esposito:1979dc,Esposito:1981dv}, ADONE $\gamma\gamma2$~\cite{Bacci:1980zs}, DCI-M3N~\cite{Cosme:1978qe}, ND~\cite{Dolinsky:1991vq}, OLYA~\cite{Kurdadze:1986tc}, and SND~\cite{Achasov:2003bv,Achasov:2009zz}.
The new measurement is in reasonable agreement with the previous experiments except for ND, which lies significantly above all others.


\begin{figure}
    \centering
    \includegraphics[type=pdf,ext=.pdf,read=.pdf, width=\linewidth]{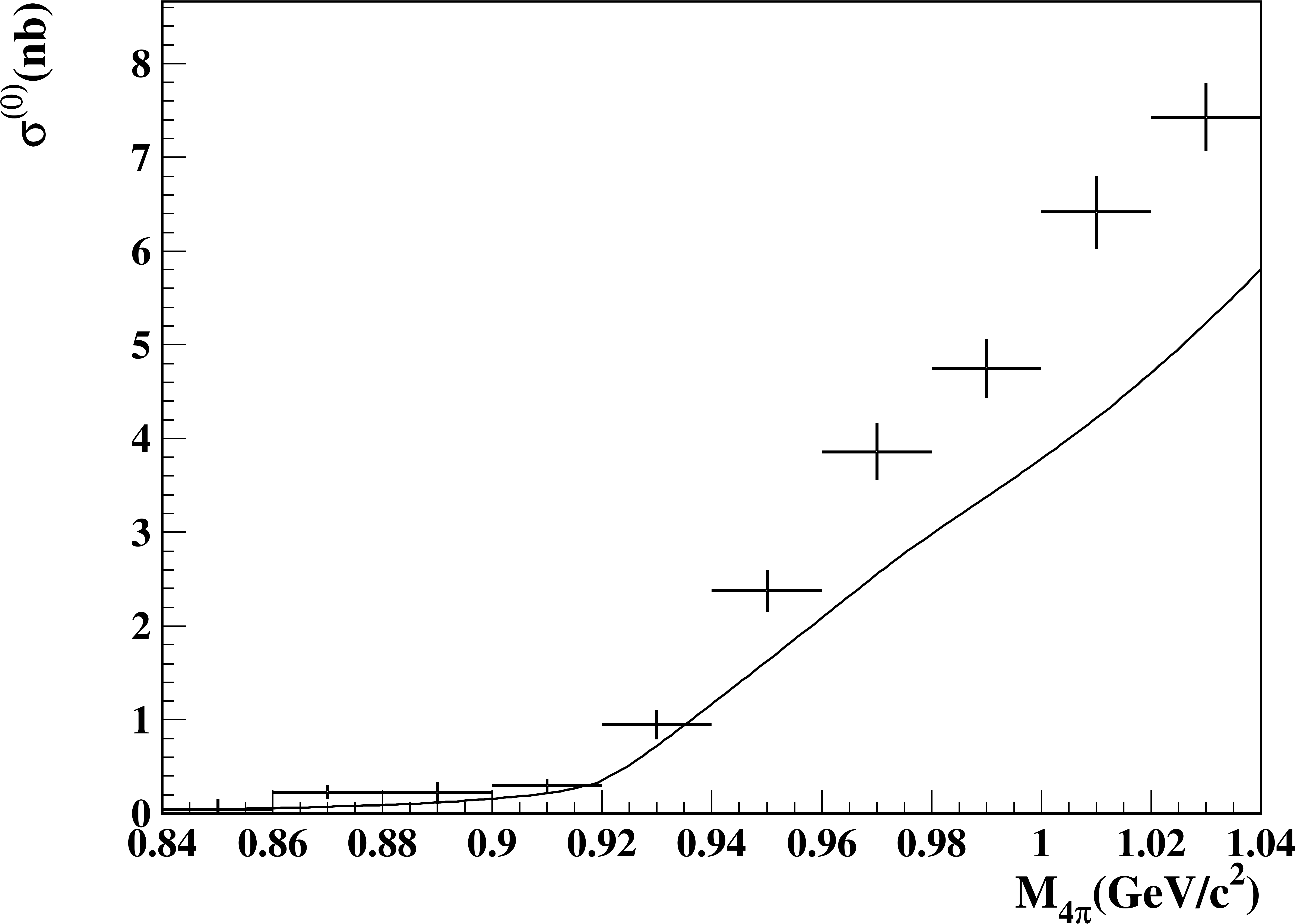}
    \caption{The low-energy part of the vacuum polarization corrected measured undressed cross section (points with statistical uncertainties) compared to the theoretical prediction (line) from Ref.~\cite{Ecker:2002cw}.}
    \label{fig:EUpred}
  \end{figure}

This cross section measurement is an important benchmark for existing theoretical calculations. In Fig.~\ref{fig:EUpred}, the prediction from chiral perturbation theory including $\omega$, $a_1$ and double $\rho$ exchange~\cite{Ecker:2002cw} is shown in comparison to data. The prediction exhibits similar behavior as the measured cross section, underestimating it slightly but especially at low energies this discrepancy is covered by the systematic uncertainties.

\subsection{Contribution to \boldmath$a_\mu$ and \boldmath$\Delta\alpha$}

The result of this analysis is of major importance for the theoretical prediction of the muon gyromagnetic anomaly $a_\mu$. Before \babar, the channel $e^+e^- \to \pi^+\pi^-2\pi^0$ was estimated to contribute approximately \SI{2.4}{\percent} of the leading order hadronic part of $a_\mu$, but the size of its uncertainty was more than one fifth of the uncertainty of all hadronic contributions combined~\cite{Davier:2003pw}. 

The theoretical prediction of $a_\mu$ relates the undressed $e^+e^-$ cross section of a given final state $X$ to the corresponding contribution to $a_\mu$ at leading order via~\cite{Jegerlehner}
\begin{equation}
a_\mu^X = \frac{1}{4\pi^3} \int_{s^X_\mathrm{min}}^\infty K_\mu(s) \cdot \frac{\sqrt{1 - \frac{4m_e^2c^4}{s}}}{1 + \frac{2m_e^2c^4}{s}} \cdot \sigma^{(0)}_{e^+e^- \to X}(s) \mathrm{d}s \text{,}
\end{equation}
where $K_\mu(s)$ is the muon kernel function and $m_e$ the electron mass~\cite{PDG}.
Integrating over the energy region $\SI{0.85}{\GeV} \le E_\mathrm{CM} \le \SI{1.8}{\GeV}$ we find 
\begin{equation}
a_\mu^{\pi^+\pi^-2\pi^0} = (\amunew \pm \amunewstaterr_\mathrm{stat} \pm \amunewsysterr_\mathrm{syst}) \times 10^{-10} \text{,}
\end{equation}
where the first uncertainty is statistical and the second systematic, giving a total relative precision of \amunewtotrelerr.

Before \babar, the world average covered the energy range $\SI{1.02}{\GeV} \le E_\mathrm{CM} \le \SI{1.8}{\GeV}$ and yielded the result\footnote{The second uncertainty corresponds to a correction of radiative effects, while the first is the combined statistical and systematic uncertainty.} $(\num{16.76} \pm \num{1.31} \pm \num{0.20}_\mathrm{rad}) \times 10^{-10}$~\cite{Davier:2003pw}, implying a total relative precision of \SI{7.9}{\percent}. 
In this region we measure $a_\mu^{\pi^+\pi^-2\pi^0} = (\amuhinew \pm \amuhinewstaterr_\mathrm{stat} \pm \amuhinewsysterr_\mathrm{syst}) \times 10^{-10}$ in agreement with the previous value. The uncertainties correspond to a total relative precision of \amuhinewtotrelerr.
Hence, the relative precision of the \babar measurement alone is a factor \num{2.5} higher than the precision of the world data set without \babar.

For comparison with theory predictions it is worthwhile extending the energy range to higher values. Hence, in the energy range $\SI{0.85}{\GeV} \le E_\mathrm{CM} \le \SI{3.0}{\GeV}$ we obtain $a_\mu^{\pi^+\pi^-2\pi^0} = (\amuwidenew \pm \amuwidenewstaterr_\mathrm{stat} \pm \amuwidenewsysterr_\mathrm{syst}) \times 10^{-10}$.

Similar to $a_\mu$, the measured undressed cross section can be used to determine this channel's contribution to the running of the fine-structure constant $\alpha$~\cite{Eidelman:1995ny}:
\begin{equation}
\alpha(q^2) = \frac{\alpha(0)}{1 - \Delta\alpha(q^2)} \text{,}
\end{equation}
where $\Delta\alpha$ is the sum of all higher order corrections and $q^2$ is the squared momentum transfer. The running of $\alpha$ is often evaluated at the $Z^0$ pole ($q^2=M^2_\mathrm{Z}c^2$). In the energy range $\SI{0.85}{\GeV} \le E_\mathrm{CM} \le \SI{1.8}{\GeV}$ the value
\begin{equation}
  \Delta\alpha^{\pi^+\pi^-2\pi^0}(M^2_\mathrm{Z}c^2) = (\alphanew \pm \alphanewstaterr_\mathrm{stat} \pm \alphanewsysterr_\mathrm{syst}) \times 10^{-4}
\end{equation}
is calculated from this measurement. For higher energies, $\SI{0.85}{\GeV} \le E_\mathrm{CM} \le \SI{3.0}{\GeV}$, we find $\Delta\alpha^{\pi^+\pi^-2\pi^0}(M^2_\mathrm{Z}c^2) = (\alphawidenew \pm \alphawidenewstaterr_\mathrm{stat} \pm \alphawidenewsysterr_\mathrm{syst}) \times 10^{-4}$.

\section{Intermediate Resonances}\label{sec:intstr}

The channel $e^+e^- \to \pi^+\pi^-2\pi^0$ is also of interest due to its internal structures. These shed light on the production process of hadrons and can probe theoretical models or provide input for the latter~\cite{Gudino:2015kra}.
In Ref.~\cite{Achasov:2003bv} it is suggested that the channel $e^+e^- \to \pi^+\pi^-2\pi^0$ is described completely by the intermediate states $a_1\pi$ and $\omega\pi^0$ in the energy range $\SI{0.98}{\GeV} < E_\mathrm{CM} < \SI{1.38}{\GeV}$. Furthermore, the authors do not observe a $\rho^0$ signal in their data, consistent with earlier measurements~\cite{Akhmetshin:1998df}. In this work, a study of the $a_1\pi$ intermediate state is undertaken but due to the large width of the $a_1$ resonance it is not possible to quantify the $a_1\pi$ contribution. The role of the $\omega\pi^0$ substructure and a possible $\rho^0$ contribution are investigated in this work over a wider energy range than in previous measurements.
A complete study of the dynamics of this process would require a partial wave analysis, preferably in combination with the channel $e^+e^- \to \pi^+\pi^-\pi^+\pi^-$. Since this is beyond the scope of this analysis, only selected intermediate states are presented here.

The efficiency as function of the mass of the sub-system is calculated using AfkQed by dividing the mass distribution after $\pi^+\pi^-2\pi^0\gamma$ selection and detector simulation by the distribution of the generated mass. Furthermore, unless stated otherwise no background subtraction is applied to data when graphing the mass distribution of a subsystem.

\begin{figure}
  \centering
  \includegraphics[width=\linewidth]{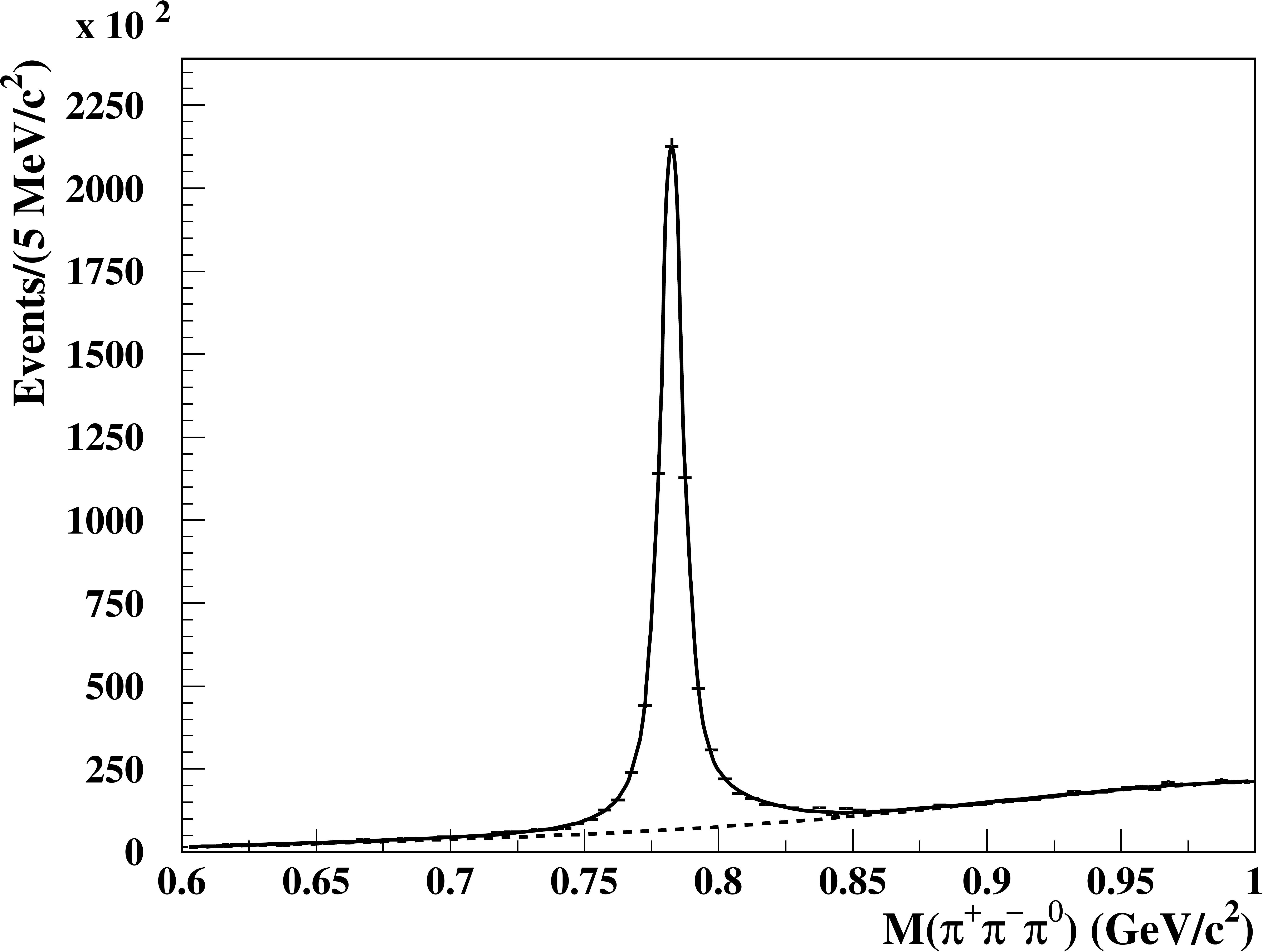}
  \caption{The measured $\omega$ data peak in the complete $M(\pi^+\pi^-2\pi^0)$ range after selection and efficiency correction.}
  \label{fig:omegaglobal}
\end{figure}

\begin{figure}
  \centering
  \includegraphics[width=\linewidth]{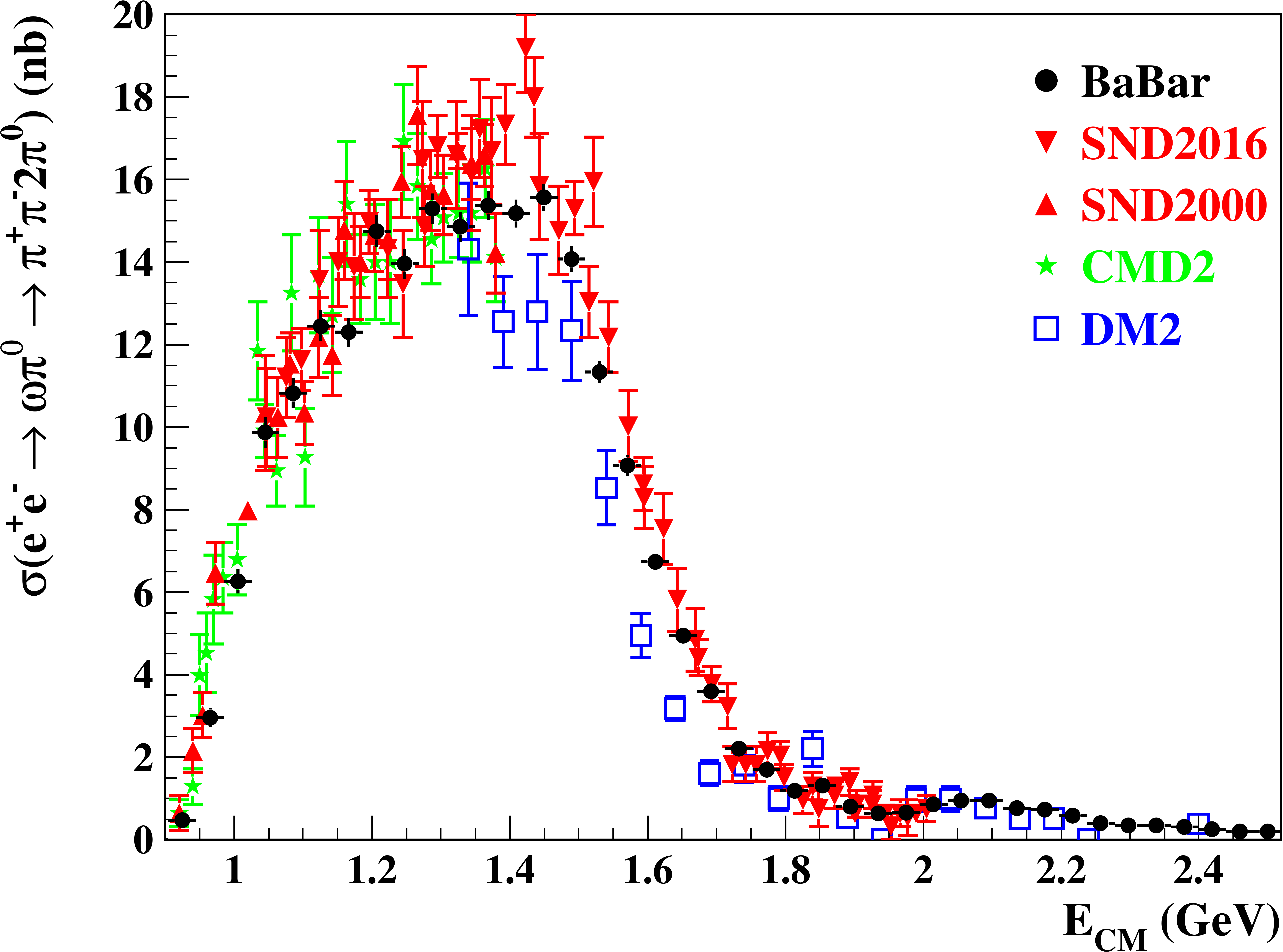}
  \colorcaption{The measured $e^+e^- \to \omega\pi^0 \to \pi^+\pi^-2\pi^0$ cross sections from different experiments~\cite{Bisello:1990du,Achasov:2000wy,Akhmetshin:2003ag,Achasov:2016zvn} as a function of $E_\mathrm{CM}$ with statistical uncertainties. Data measured in other decays than $\omega \to \pi^+\pi^-\pi^0$ is scaled by the appropriate branching ratio.}
  \label{fig:omerate}
\end{figure}

One important intermediate state is given by the channel $e^+e^- \to \omega\pi^0\gamma \to \pi^+\pi^-2\pi^0\gamma$ with $\mathcal{B}(\omega \to \pi^+\pi^-\pi^0) = \num{0.892} \pm \num{0.007}$~\cite{PDG}. Fitting a Voigt profile plus a normal distribution (for the radiative tail) to the efficiency corrected $M(\pi^+\pi^-\pi^0)$ distribution, as shown in Fig.~\ref{fig:omegaglobal}, results in an $\omega\pi^0$ production fraction of \SI[parse-numbers=false]{(32.1 \pm 0.2_\mathrm{stat} \pm 2.6_\mathrm{syst})}{\percent} over the full invariant mass range. The systematic uncertainty is determined as the difference from an alternative fit function.
The same fitting procedure is applied in narrow slices of the invariant mass $M(\pi^+\pi^-2\pi^0)$. The resulting number of events is divided by the ISR-luminosity in each mass region, yielding the cross section $\sigma(e^+e^- \to \omega\pi^0\gamma \to \pi^+\pi^-2\pi^0\gamma)$ as a function of the CM-energy of the hadronic system listed in Table~\ref{tab:csomega} and shown in Fig.~\ref{fig:omerate} in comparison to existing data~\cite{Bisello:1990du,Achasov:2000wy,Akhmetshin:2003ag,Achasov:2016zvn}. In this case, possible background processes are removed by the fit function.
 The $\omega\pi^0$ production fraction dominates at low masses, then decreases rapidly, such that it is on the level of \SI{10}{\percent} already at $M(\pi^+\pi^-2\pi^0) \approx \SI{1.8}{\GeVpercsq}$, decreasing further towards higher masses.

  \LTcapwidth=\linewidth
\begin{table}
\caption{The measured $e^+e^- \to \omega \pi^0 \to \pi^+\pi^-\pi^0\pi^0$ cross section with statistical uncertainties. The relative systematic uncertainty amounts to \SI{10}{\percent}.}
\begin{tabular}{c r}
\multicolumn{1}{c}{$E_\mathrm{CM} (\si{\GeV})$} & \multicolumn{1}{c}{$\sigma (\si{\nano\barn})$} \\
\hline
0.924 & 0.48 $\pm$ 0.08 \\
0.965 & 2.96 $\pm$ 0.23 \\
1.005 & 6.26 $\pm$ 0.30 \\
1.045 & 9.87 $\pm$ 0.37 \\
1.086 & 10.82 $\pm$ 0.37 \\
1.126 & 12.45 $\pm$ 0.38 \\
1.167 & 12.30 $\pm$ 0.36 \\
1.207 & 14.75 $\pm$ 0.38 \\
1.247 & 13.95 $\pm$ 0.36 \\
1.288 & 15.30 $\pm$ 0.37 \\
1.328 & 14.85 $\pm$ 0.35 \\
1.369 & 15.37 $\pm$ 0.35 \\
1.409 & 15.19 $\pm$ 0.34 \\
1.449 & 15.57 $\pm$ 0.34 \\
1.490 & 14.22 $\pm$ 0.30 \\
1.530 & 11.52 $\pm$ 0.26 \\
1.571 & 9.05 $\pm$ 0.25 \\
1.611 & 6.66 $\pm$ 0.20 \\
1.652 & 4.94 $\pm$ 0.20 \\
1.692 & 3.52 $\pm$ 0.14 \\
1.732 & 2.21 $\pm$ 0.11 \\
1.773 & 1.68 $\pm$ 0.09 \\
1.813 & 1.19 $\pm$ 0.08 \\
1.854 & 1.30 $\pm$ 0.08 \\
1.894 & 0.80 $\pm$ 0.07 \\
1.934 & 0.63 $\pm$ 0.06 \\
1.975 & 0.65 $\pm$ 0.06 \\
2.015 & 0.85 $\pm$ 0.06 \\
2.056 & 0.94 $\pm$ 0.07 \\
2.096 & 0.95 $\pm$ 0.07 \\
2.136 & 0.77 $\pm$ 0.06 \\
2.177 & 0.73 $\pm$ 0.05 \\
2.217 & 0.58 $\pm$ 0.05 \\
2.258 & 0.40 $\pm$ 0.04 \\
2.298 & 0.34 $\pm$ 0.04 \\
2.338 & 0.35 $\pm$ 0.04 \\
2.379 & 0.31 $\pm$ 0.03 \\
2.419 & 0.25 $\pm$ 0.03 \\
2.460 & 0.20 $\pm$ 0.03 \\
2.500 & 0.20 $\pm$ 0.03 \\
\hline
\end{tabular}
\label{tab:csomega}
\end{table}
  
\begin{figure}
  \centering
  \includegraphics[trim = 0 0 0 38mm, clip, width=\linewidth]{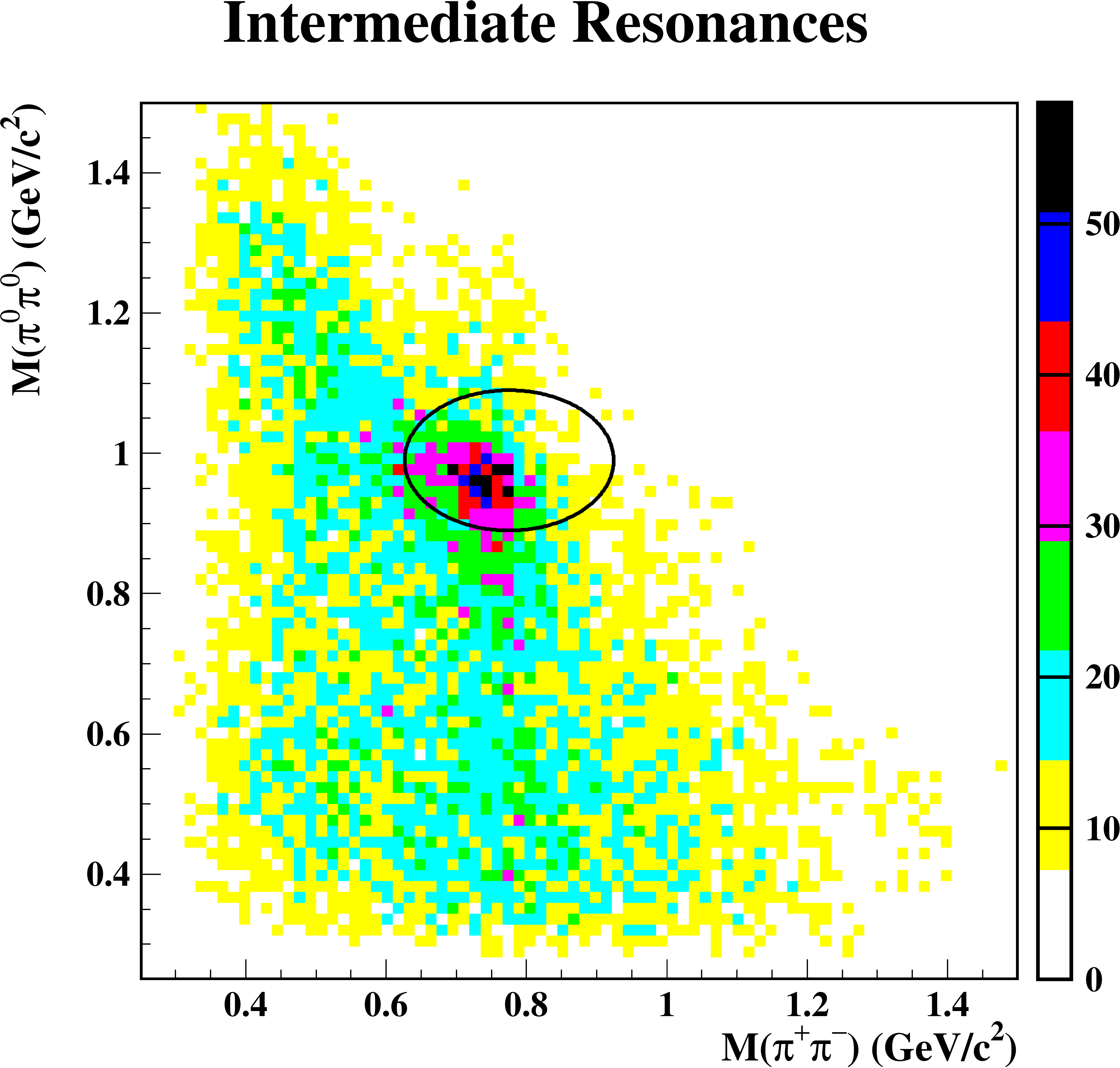}
  \colorcaption{The $M(\pi^0\pi^0)$ vs. $M(\pi^+\pi^-)$ 2D plot of data in the invariant mass interval $\SI{1.7}{\GeVpercsq} < M(\pi^+\pi^-2\pi^0) < \SI{2.3}{\GeVpercsq}$ after selection without efficiency correction and background subtraction. The black ellipse indicates the region used to select $\rho^0f_0$ events.}
  \label{fig:m2pim2pi0scat}
\end{figure}

\begin{figure}
  \centering
  \includegraphics[width=\linewidth]{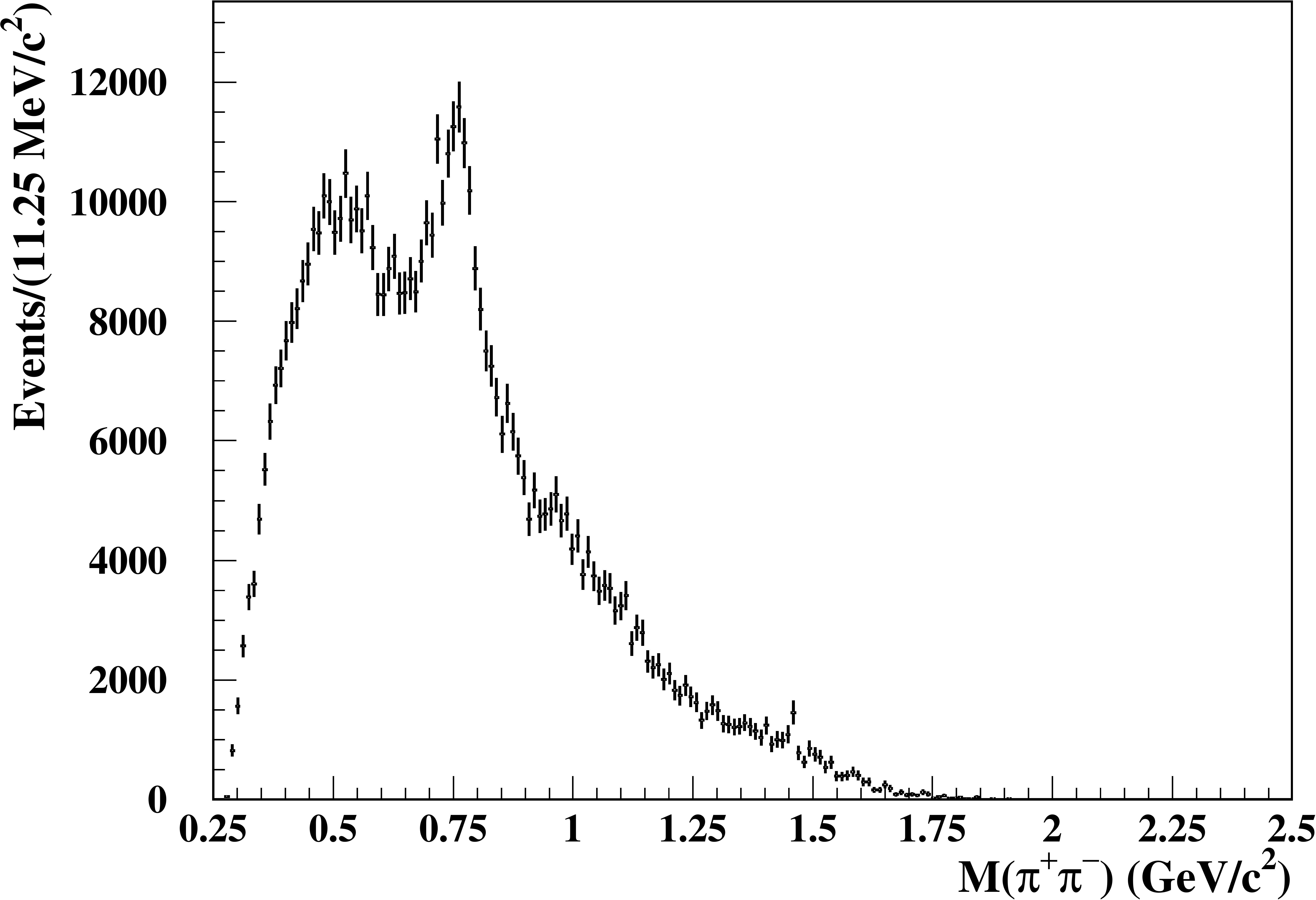}
  \caption{The $M(\pi^+\pi^-)$ distribution in the invariant mass interval $\SI{1.7}{\GeVpercsq} < M(\pi^+\pi^-2\pi^0) < \SI{2.3}{\GeVpercsq}$ for data after selection and efficiency correction.}
  \label{fig:mpippim}
\end{figure}

\begin{figure}
  \centering
  \includegraphics[width=\linewidth]{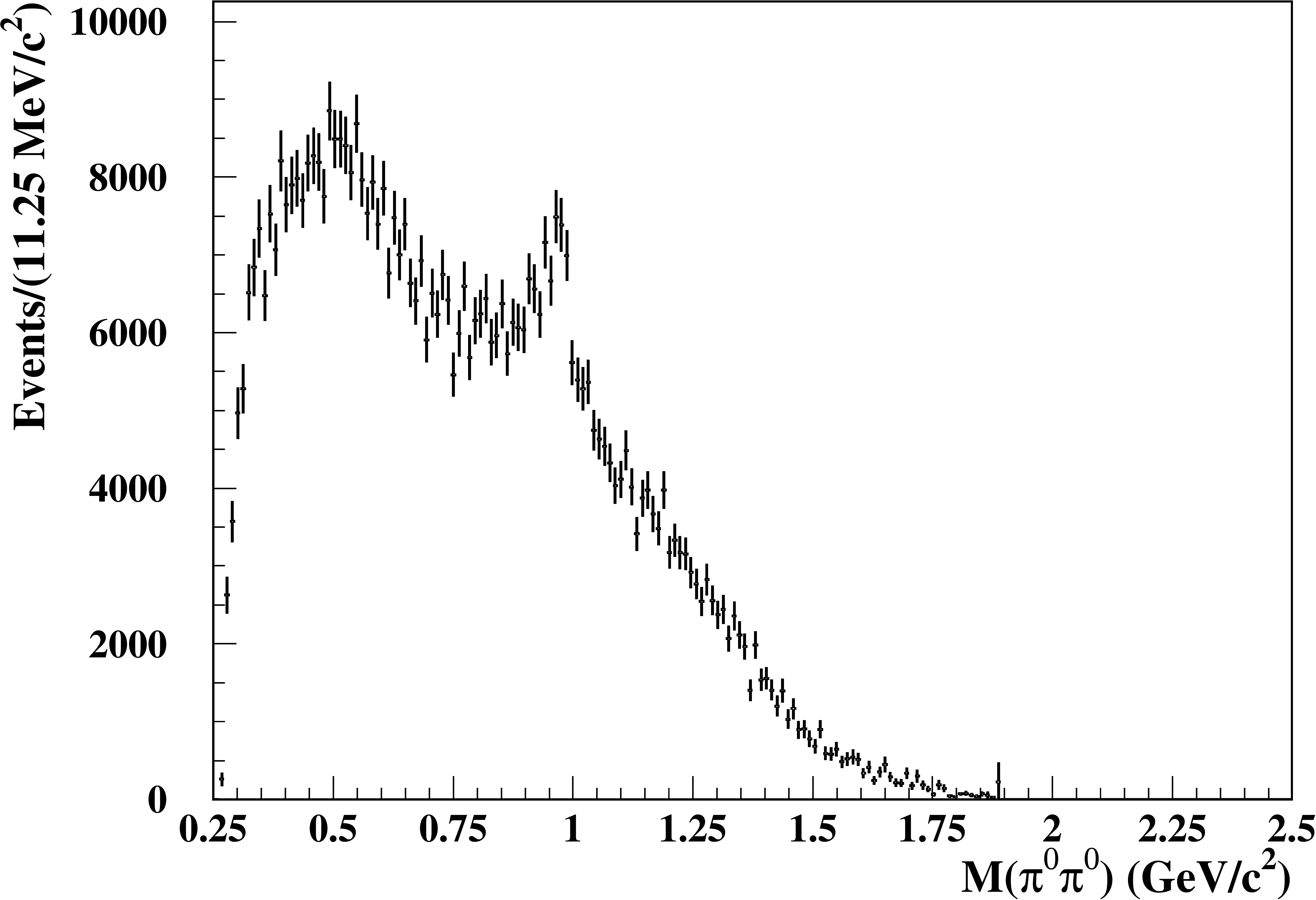}
  \caption{The $M(\pi^0\pi^0)$ distribution in the invariant mass interval $\SI{1.7}{\GeVpercsq} < M(\pi^+\pi^-2\pi^0) < \SI{2.3}{\GeVpercsq}$ for data after selection and efficiency correction.}
  \label{fig:mpi0pi0}
\end{figure}

\begin{figure}
  \centering
  \includegraphics[width=\linewidth]{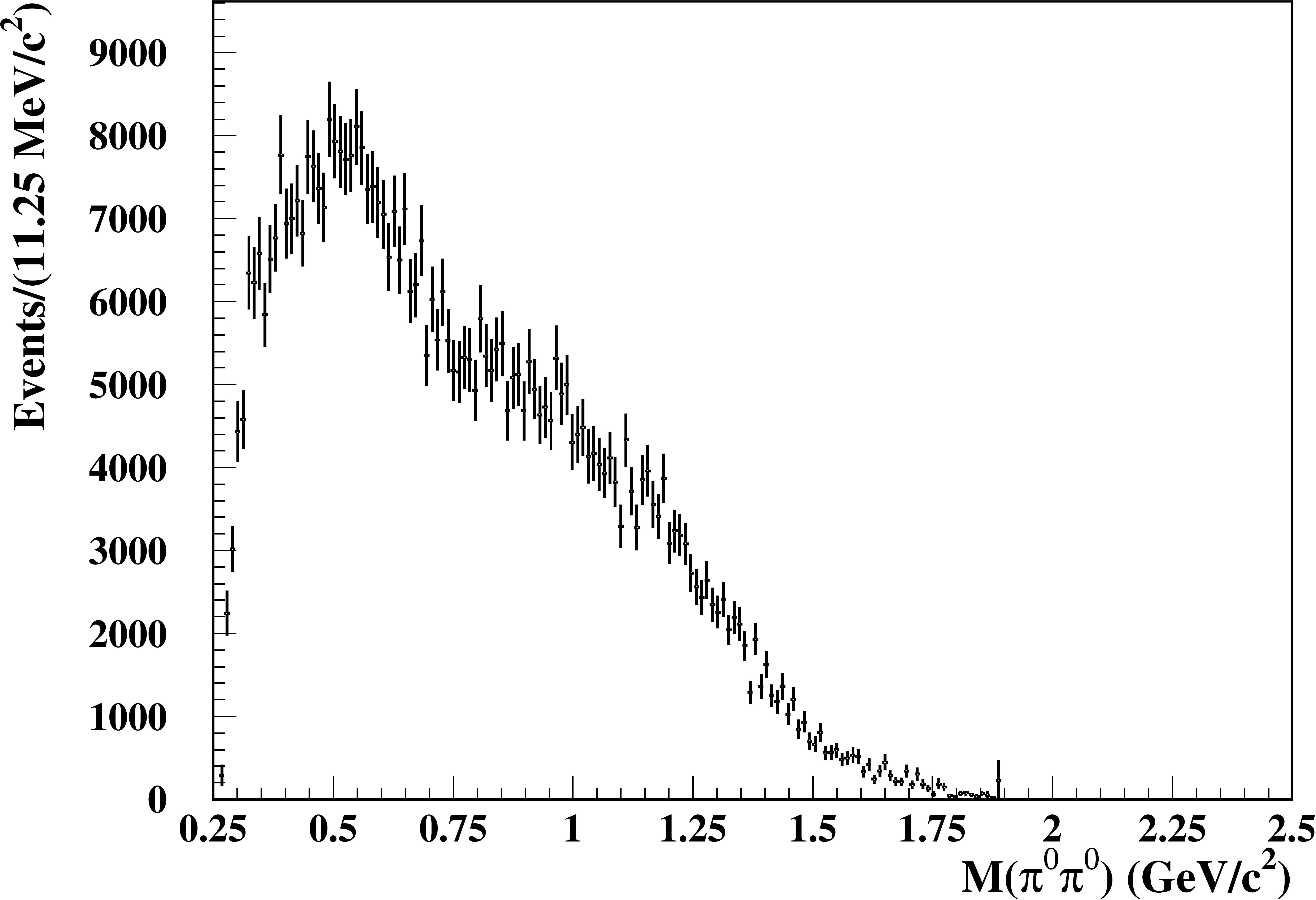}
  \caption{The $M(\pi^0\pi^0)$ distribution excluding the $\rho^0$ mass range in $M(\pi^+\pi^-)$ in the invariant mass interval $\SI{1.7}{\GeVpercsq} < M(\pi^+\pi^-2\pi^0) < \SI{2.3}{\GeVpercsq}$ for data after selection and efficiency correction.}
  \label{fig:mpi0pi0_norho}
\end{figure}

Figure~\ref{fig:m2pim2pi0scat} shows the 2D plot of the $\pi^+\pi^-$ mass vs. the $\pi^0\pi^0$ mass in the range $\SI{1.7}{\GeVpercsq} < M(\pi^+\pi^-2\pi^0) < \SI{2.3}{\GeVpercsq}$, which is chosen to achieve the best prominence of observed structures. In this mass region, the distribution exhibits an excess of events around $M(\pi^+\pi^-) \approx \SI{0.77}{\GeVpercsq}$ and $M(\pi^0\pi^0) \approx \SI{1.0}{\GeVpercsq}$.
Investigating this structure in the efficiency corrected one-dimensional distribution in $M(\pi^+\pi^-)$, Fig.~\ref{fig:mpippim}, shows a substantial peak near the $\rho^0$ mass.
Figure~\ref{fig:mpi0pi0} shows that the peak in the $M(\pi^0\pi^0)$ distribution is around the $f_0(980)$ mass with a sharp edge just above the peak. Moreover, this peak vanishes when rejecting events from the $\rho^0$ region in $M(\pi^+\pi^-)$ as observed in Fig.~\ref{fig:mpi0pi0_norho}, implying production exclusively in combination with a $\rho^0$.

\begin{figure}
  \centering
  \includegraphics[trim = 0 0 0 38mm, clip, width=\linewidth]{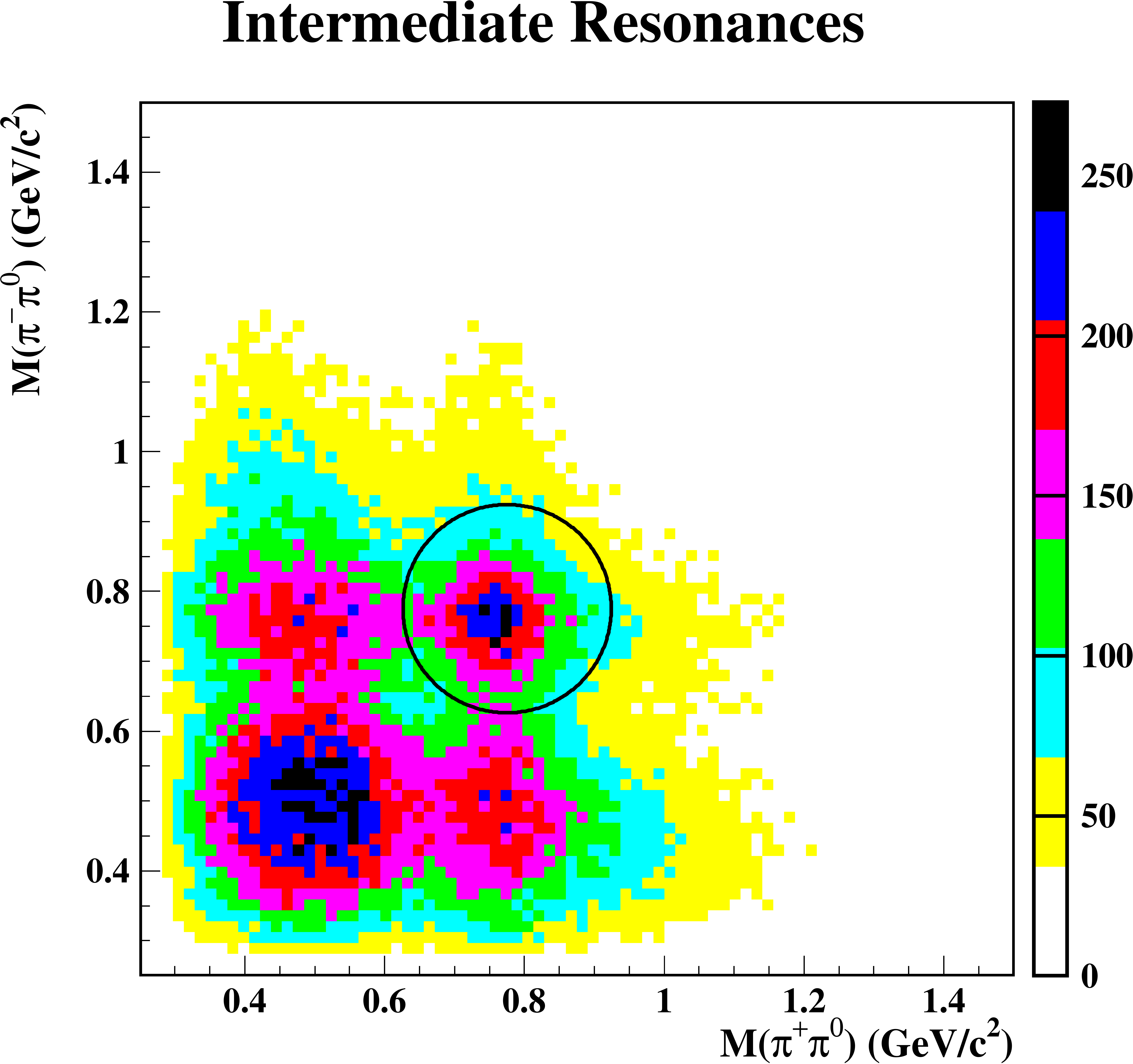}
  \colorcaption{The $M(\pi^+\pi^0)$ vs. $M(\pi^-\pi^0)$ 2D plot of data after selection without efficiency correction and background subtraction. The black circle indicates the region used to select $\rho^+\rho^-$ events.}
  \label{fig:mpimpi0mpippi0scat}
\end{figure}

\begin{figure}
  \centering
  \includegraphics[width=\linewidth]{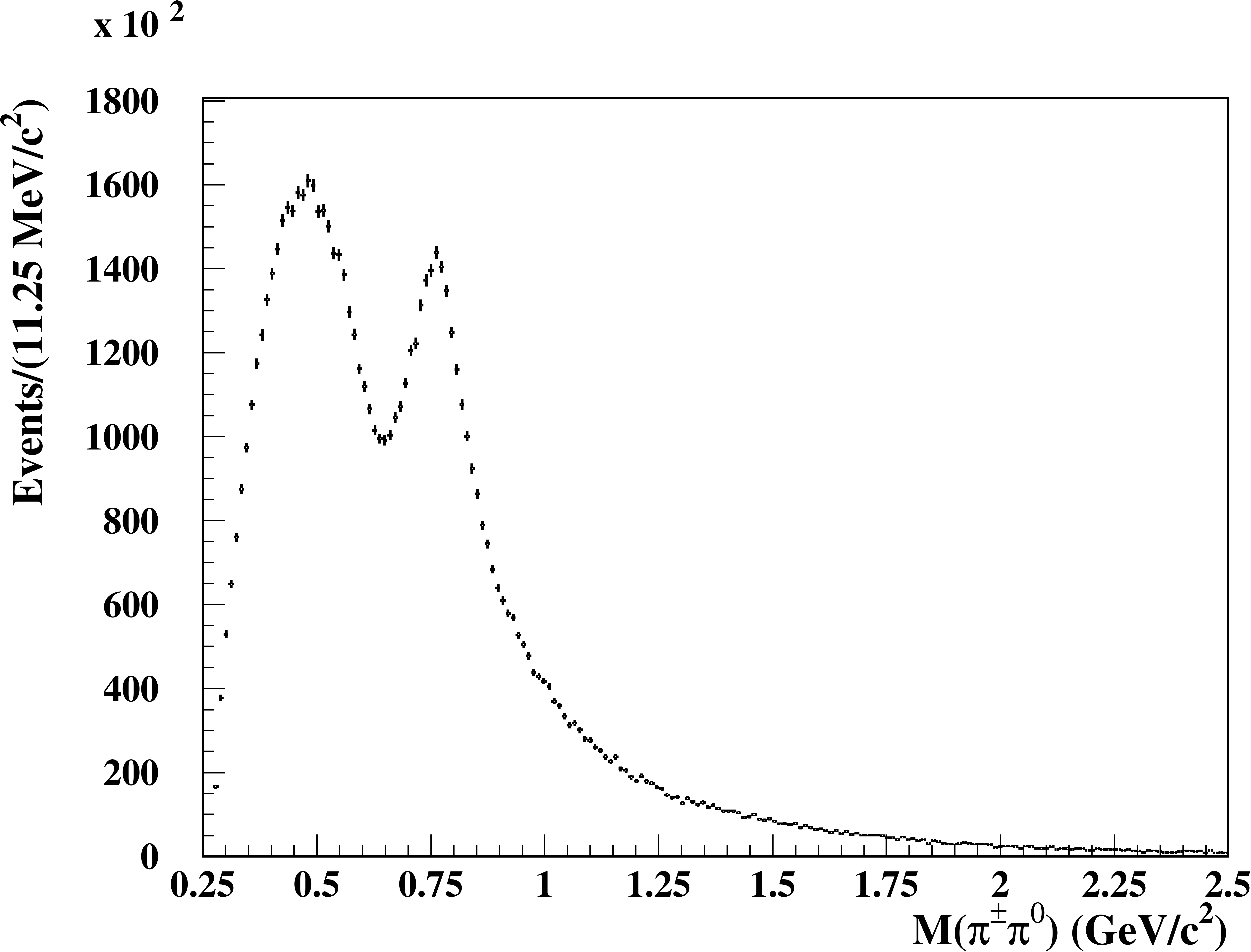}
  \caption{The $M(\pi^\pm\pi^0)$ distribution in data after selection and efficiency correction.}
  \label{fig:mpipmpi0}
\end{figure}

In the other two-pion combination, the masses $M(\pi^\pm\pi^0)$ are studied, whose 2D plot is shown in Fig.\ref{fig:mpimpi0mpippi0scat}.
Correlated $\rho^+\rho^-$ production is visible as a peak around the $\rho^+\rho^-$ mass-crossing and has not been observed before. In the one-dimensional $M(\pi^\pm\pi^0)$ distribution, Fig.~\ref{fig:mpipmpi0}, a large $\rho^\pm$ peak is observed in data.

\begin{figure}
  \centering
  \includegraphics[width=\linewidth]{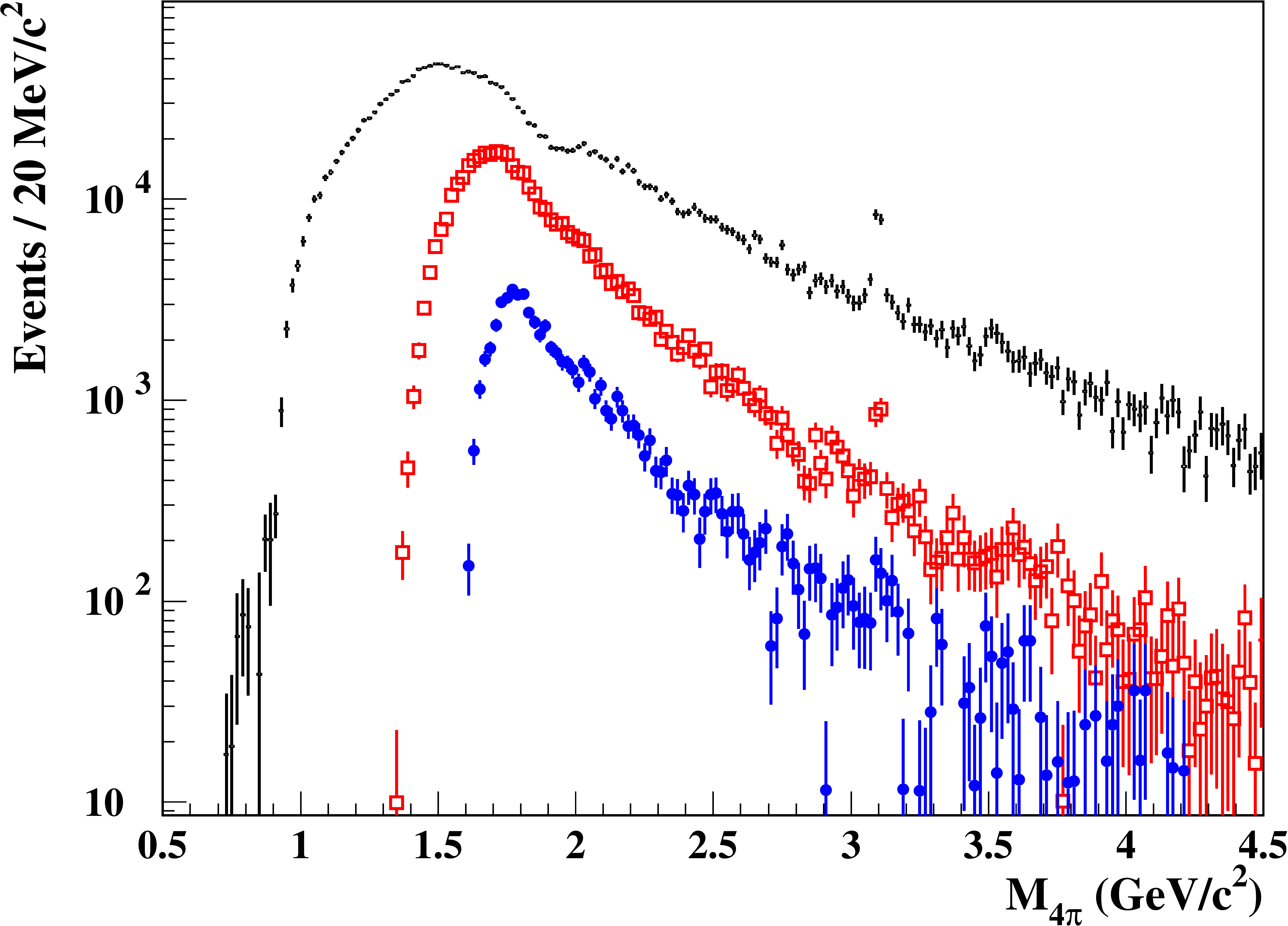}
  \colorcaption{The mass spectra of $\pi^+\pi^-2\pi^0$ events (after background subtraction and efficiency correction) from data in the $\rho^+\rho^-$ (red squares), $\rho^0f_0$ regions (blue circles), and in the full range (black points).}
  \label{fig:cs_intres}
\end{figure}

If background processes are subtracted using simulation for continuum and ISR processes (as outlined in Sec.~\ref{sec:bkg}) and normalization to efficiency is applied, the $e^+e^- \to \pi^+\pi^-2\pi^0$ mass spectrum can be obtained specifically for resonance regions.
Restricting the two-$\pi^0$ mass to the $f_0$ region $\SI{0.89}{\GeVpercsq} < M(\pi^0\pi^0) < \SI{1.09}{\GeVpercsq}$ and the $\pi^+\pi^-$ mass to the $\rho^0$ region $\SI{0.63}{\GeVpercsq} < M(\pi^+\pi^-) < \SI{0.92}{\GeVpercsq}$, as indicated by the black ellipse in Fig.~\ref{fig:m2pim2pi0scat}, results in the mass spectrum shown as the blue circles in Fig.~\ref{fig:cs_intres}. 
Similarly, restricting the $\pi^\pm\pi^0$ masses to the $\rho^\pm$ region $\SI{0.63}{\GeVpercsq} < M(\pi^\pm\pi^0) < \SI{0.92}{\GeVpercsq}$, as indicated by the black circle in Fig.~\ref{fig:mpimpi0mpippi0scat}, results in the mass spectrum shown as the red squares in Fig.~\ref{fig:cs_intres}. 
Although backgrounds from processes besides the signal $e^+e^- \to \pi^+\pi^-2\pi^0$ are subtracted, the mass spectra in both resonance regions still include a sizable fraction of events not produced via the intermediate states $\rho^0f_0$ or $\rho^+\rho^-$, respectively. Nonetheless, a peaking structure is visible especially in the $\rho^0f_0$ distribution.

\section{\jpsi Branching Fraction}\label{sec:jpsi}

\begin{figure}
  \centering
  \includegraphics[trim = 0 0 0 38mm, clip, width=\linewidth]{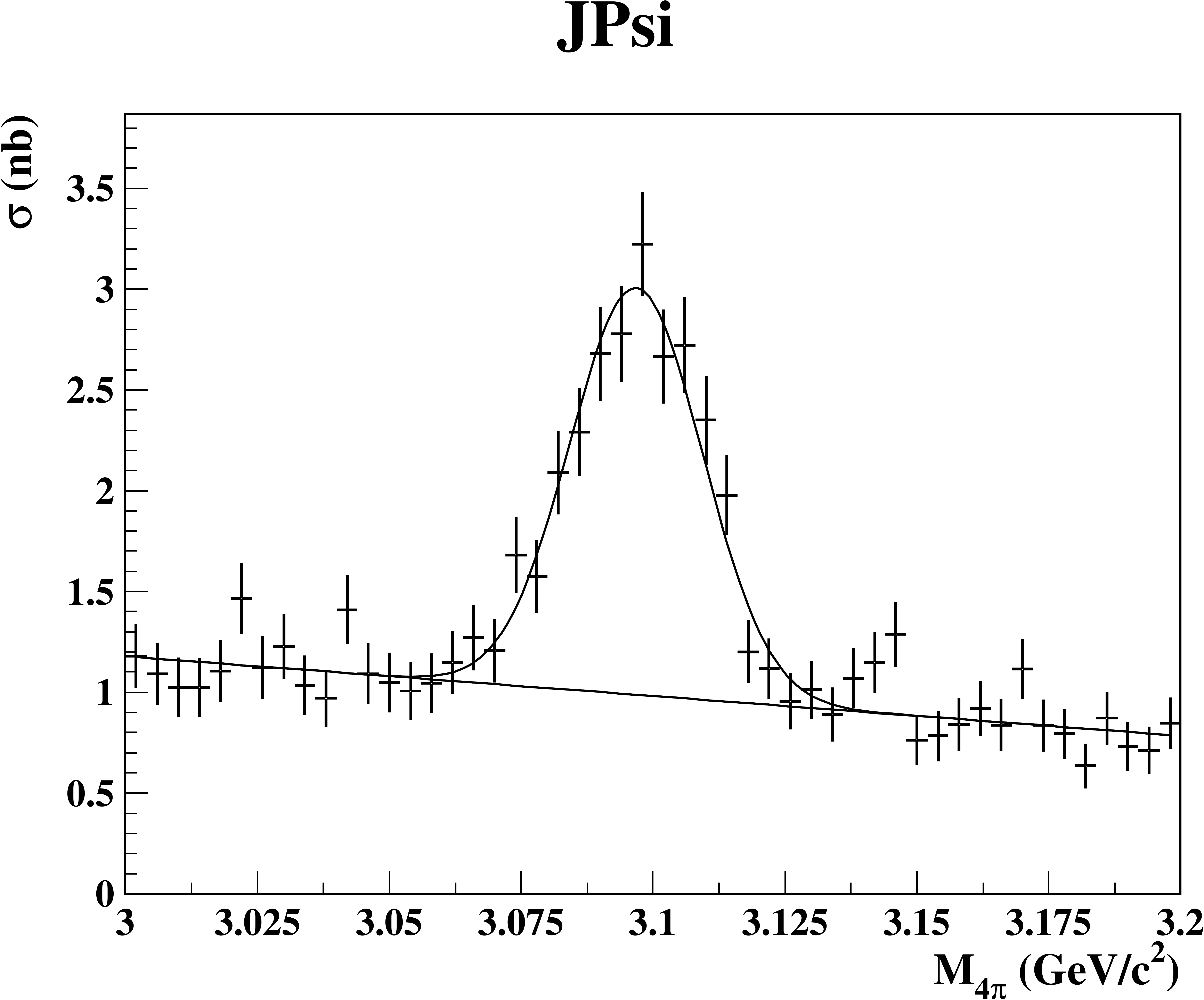}
  \caption{The measured $J/\psi$ peak in the $\pi^+\pi^-2\pi^0$ cross section without background subtraction.}
  \label{fig:jpsi}
\end{figure}

The \jpsi peak in the $\pi^+\pi^-2\pi^0$ cross section is used to determine the branching ratio of $\jpsi \to \pi^+\pi^-2\pi^0$. For this purpose, the number of \jpsi events in the channel $e^+e^- \to \pi^+\pi^-2\pi^0$ normalized to luminosity is obtained from data using the Gaussian fit shown in Fig.~\ref{fig:jpsi} and is corrected for non-normality of the mass resolution. A linear parametrization is employed for the background, which is dominated by non-resonant $e^+e^- \to \pi^+\pi^-2\pi^0$ production.
From the fit, the product of the integrated \jpsi cross section and the branching fraction $\jpsi \to \pi^+\pi^-2\pi^0$ is determined:
\begin{equation}
  \mathcal{B}_{J/\psi \rightarrow \pi^+\pi^-2\pi^{0}}\sigma^{\jpsi}_\text{int} = \SI[parse-numbers=false]{(68 \pm 4_\mathrm{stat} \pm 5_\mathrm{fit} )}{\nano\barn \MeVpercsq} \text{.}
\end{equation}
From the integrated cross section of a resonance the following relation for calculating the branching fraction is derived~\cite{nagashima1} (with $M_{J/\psi} = \SI[parse-numbers=false]{(3096.900 \pm 0.006)}{\MeVpercsq}$~\cite{PDG}):
\begin{equation}
\begin{aligned}
  \mathcal{B}_{J/\psi \rightarrow \pi^+\pi^-2\pi^{0}} \Gamma^{J/\psi}_{ee} &= \frac{N(J/\psi \rightarrow \pi^+\pi^-2\pi^{0}) \cdot M^2_{J/\psi}c^4}{6 \pi^2 \hbar^2 c^2 \cdot \mathrm{d}\mathcal{L}/\mathrm{d}E \cdot \varepsilon} \\
&= \SI[parse-numbers=false]{(28.3 \pm 1.7_\mathrm{stat} \pm 2.1_\mathrm{syst})}{\electronvolt} \text{,}
\end{aligned}
\end{equation}
where $\varepsilon$ is the detection efficiency and the input uncertainty is negligible. If this value is divided by $\Gamma^{J/\psi}_{ee} = \SI[parse-numbers=false]{(5.55 \pm 0.14)}{\keV}$~\cite{PDG}, the branching fraction follows:
\begin{equation}
  \mathcal{B}_{J/\psi \rightarrow \pi^+\pi^-2\pi^{0}} = (\num{5.1} \pm \num{0.3}_\mathrm{stat} \pm \num{0.4}_\mathrm{syst} \pm \num{0.1}_\mathrm{input}) \times 10^{-3} \text{,}
\end{equation}
where the input uncertainty is the propagation of the uncertainties of $M^2_{J/\psi}$, $\Gamma^{J/\psi}_{ee}$, and $\hbar c$. The systematic uncertainty is determined by the systematic uncertainty of the general analysis with the exception of the background subtraction. In this study, the background is subtracted via the fit function and hence its systematic uncertainty is included in the model error, which is determined by fitting several peak and background shapes to data.


\section{Summary and Conclusions}
\label{sec:summary}

In this study, the cross section $e^+e^- \to \pi^+\pi^-2\pi^0$ is measured with unprecedented precision. At large invariant masses $M(\pi^+\pi^-2\pi^0) > \SI{3.2}{\GeVpercsq}$, a systematic precision of \systhigh\ is reached, while in the region $\SI{2.7}{\GeVpercsq} < M(\pi^+\pi^-2\pi^0) < \SI{3.2}{\GeVpercsq}$ it is \systjpsi. In the peak region $\threshul < M(\pi^+\pi^-2\pi^0) < \SI{2.7}{\GeVpercsq}$ a relative systematic uncertainty of \systpeak\ is achieved.

This measurement is subsequently used to calculate the channel's contribution to $a_\mu$ in the energy range $\SI{0.85}{\GeV} \le E_\mathrm{CM} \le \SI{1.8}{\GeV}$:
\begin{equation}
a_\mu(\pi^+\pi^-2\pi^0) = (\amunew \pm \amunewstaterr_\mathrm{stat} \pm \amunewsysterr_\mathrm{syst}) \times 10^{-10} \text{.}
\end{equation}
For $\SI{0.85}{\GeV} \le E_\mathrm{CM} \le \SI{3.0}{\GeV}$ we obtain
\begin{equation}
a_\mu^{\pi^+\pi^-2\pi^0} = (\amuwidenew \pm \amuwidenewstaterr_\mathrm{stat} \pm \amuwidenewsysterr_\mathrm{syst}) \times 10^{-10} \text{.}
\end{equation}

Furthermore, intermediate structures from the channels $\rho^0f_0$ and $\rho^+\rho^-$ are seen. The contribution produced via $\omega\pi^0$ is studied and the cross section measured. The branching fraction $\jpsi \to \pi^+\pi^-2\pi^0$ is determined. For a deeper understanding of the production mechanism, a partial wave analysis in combination with the process $e^+e^- \to \pi^+\pi^-\pi^+\pi^-$~\cite{Lees:2012cr} is necessary.

\setcounter{footnote}{0}

\section{Acknowledgments}
\label{sec:Acknowledgments}
We are grateful for the 
extraordinary contributions of our \pep2\ colleagues in
achieving the excellent luminosity and machine conditions
that have made this work possible.
The success of this project also relies critically on the 
expertise and dedication of the computing organizations that 
support \babar.
The collaborating institutions wish to thank 
SLAC for its support and the kind hospitality extended to them. 
This work is supported by the
US Department of Energy
and National Science Foundation, the
Natural Sciences and Engineering Research Council (Canada),
the Commissariat \`a l'Energie Atomique and
Institut National de Physique Nucl\'eaire et de Physique des Particules
(France), the
Bundesministerium f\"ur Bildung und Forschung and
Deutsche Forschungsgemeinschaft
(Germany), the
Istituto Nazionale di Fisica Nucleare (Italy),
the Foundation for Fundamental Research on Matter (The Netherlands),
the Research Council of Norway, the
Ministry of Education and Science of the Russian Federation, 
Ministerio de Econom\'{\i}a y Competitividad (Spain), the
Science and Technology Facilities Council (United Kingdom),
and the Binational Science Foundation (U.S.-Israel).
Individuals have received support from 
the Marie-Curie IEF program (European Union) and the A. P. Sloan Foundation (USA). 
The United States Government retains and the publisher, by accepting the article for publication, acknowledges that the United States Government retains a non-exclusive, paid-up, irrevocable, world-wide license to publish or reproduce the published form of this manuscript, or allow others to do so, for United States Government purposes.

\bibliography{2pi2pi0-pd}

\end{document}